\documentclass[11pt,fleqn]{article}%
\usepackage[scale={.8,.86}]{geometry}
\usepackage[latin1]{inputenc}
\usepackage{axodraw,subfigure,caption2}%,wrapfig
\usepackage{amsmath,amssymb,amsfonts,bbm,mathrsfs,dcolumn}
\usepackage[hypertex]{hyperref}
\usepackage{color}
\usepackage{psfrag}
\usepackage{epsfig}

\newcolumntype{d}{D{.}{.}{1}}

\newlength{\textlength}
\newlength{\overlinelength}
\newcommand{\ovl}[2][.55]{\settowidth{\textlength}{$#2$}
  \setlength{\overlinelength}{0.1pt}
  \addtolength{\overlinelength}{0.75\textlength}
  \makebox[\textlength][s]{$#2$} \hspace{-.55\textlength}
  \hspace{-\overlinelength}\hspace{#1\overlinelength}
  \overline{\makebox[\overlinelength][s]{\vphantom{$#2$}}}
  \hspace{-#1\overlinelength}\hspace{.55\textlength}}

\DeclareMathOperator{\diag}{diag}
\DeclareMathOperator{\re}{Re}
\DeclareMathOperator{\im}{Im}

\newcommand{\abs}[1]{\left| #1\right|}
\newcommand{\braket}[3]{\left\langle #1\!\left| #2\right|\! #3\right\rangle} % < | >

\newcommand{\VEV}[1]{\left\langle #1\right\rangle}

\newcommand{\tev}{\,\mbox{TeV}}
\newcommand{\gev}{\,\mbox{GeV}}
\newcommand{\mev}{\,\mbox{MeV}}

\begin{document}
\noindent
TTP09-10  \hfill April 2009 \\ 
SFB/CPP-09-31 
\vspace*{4mm}
\begin{center}
  {\Large\bfseries Probing Yukawa Unification with K and B Mixing}
  \\ \vspace*{3mm}
  {\large St\'ephanie Trine$^1$, Susanne Westhoff$^1$, and S\"oren Wiesenfeldt$^{1,2}$}
  \\ \vspace*{1mm}
  {\normalsize {\slshape $^1$ Institut f\"ur Theoretische Teilchenphysik,
      Universit\"at Karlsruhe, 76128 Karlsruhe, Germany }}
      \\
  {\normalsize {\slshape $^2$ Helmholtz Association, Anna-Louisa-Karsch-Stra{\ss}e 2, 10178 Berlin, Germany }}
\end{center}
\vspace*{0mm}
\begin{abstract}
\noindent
  We consider corrections to the unification of down-quark and
    charged-lepton Yukawa couplings in supersymmetric GUTs, which links
    the large $\nu_\tau-\nu_\mu$ mixing angle to $b\to s$ transitions.
    These corrections generically occur in simple grand-unified models
    with small Higgs representations and affect $s\to d$ and $b\to d$
    transitions via the mixing of the corresponding right-handed
    superpartners.  On the basis of a specific SUSY-SO(10) model, we
    analyze the constraints from $K-\ovl{K}$ and $B_d-\ovl{B_d}$
    mixing on the additional $\tilde{d}_R-\tilde{s}_R$ rotation angle
    $\theta$.  We find that $\epsilon_K$ already sets a stringent
    bound on $\theta$, $\theta^{\rm
      max}\sim\mathcal{O}\left(1^\circ\right)$, indicating a very
    specific flavor structure of the correction operators.  The impact
    of the large neutrino mixings on the unitarity triangle analysis
    is also briefly discussed, as well as their ability to account for
    the sizeable CP-violating phase observed recently in $B_s\to J/
    \psi\phi$~decays.
\end{abstract}

%%%%%%%%%%%%%%%%%%%%%%%%%%%%%%%%%%%%%%%%%%%%%%%%%%%%%%%%%%%%%%%%%%%%%%%%
\section{Introduction}
%%%%%%%%%%%%%%%%%%%%%%%%%%%%%%%%%%%%%%%%%%%%%%%%%%%%%%%%%%%%%%%%%%%%%%%%

The start of the LHC at CERN will enable us to study TeV-scale
physics directly for the first time.  Most importantly, we will eventually probe
the mechanism of electroweak symmetry breaking; moreover, we will be
able to test various scenarios for new physics beyond the standard model
(SM), the leading candidate of which
is arguably supersymmetry (SUSY).  The presence of supersymmetry at the
TeV-scale eliminates the quadratically-divergent loop contributions to
the Higgs mass and thereby stabilizes the electroweak scale against the
scales of more fundamental physics.  In addition, TeV-scale SUSY models
provide an attractive mechanism for electroweak symmetry breaking and an appealing candidate for cold dark matter.  Furthermore, they offer a
compelling outline for the unification of all matter and interactions,
the first step of which is grand unification \cite{Pati:1973uk}.

The near unification of the SM gauge couplings within the minimal
supersymmetric standard model (MSSM) at $M_\text{GUT} \simeq 2 \cdot
10^{16}$ GeV, with the MSSM being valid above the TeV-scale, suggests that
the standard-model group, $\text{G}_\text{SM} = \text{SU(3)}_C \times
\text{SU(2)}_L \times \text{U(1)}_Y$, is embedded into a simple gauge
group at this scale, such as SU(5) \cite{Georgi:1974sy} or SO(10)
\cite{so10}.
SO(10) is arguably the most natural GUT group: both the SM gauge and
matter fields are unified, introducing only one additional matter
particle, the right-handed neutrino.\footnote{Strictly speaking, it is
  the left-handed neutrino singlet.} It is an anomaly-free theory and
therefore explains the intricate cancellation of the anomalies in the
standard model \cite{Georgi:1972bb}.  Moreover, it contains $B-L$ as a
local symmetry, where $B$ and $L$ are baryon and lepton number,
respectively; the breaking of $B-L$ provides light neutrino masses via
the seesaw mechanism.
Remarkably, $M_\text{GUT}$ is of the right order of magnitude to
generate neutrino masses in the sub-eV range.  Hence, the neutrino
masses are linked to the breaking of the GUT symmetry
\cite{Minkowski:1977sc}.

Flavor experiments, though not able to access the TeV scale directly,
have put strong constraints on the MSSM parameters.  Due to the lack of
deviation with respect to the SM, we expect the new sources
of flavor mixing and CP violation to be very limited for SUSY particles around the weak scale.
As formulated by the concept of minimal flavor violation
\cite{minimal-flavor,Buras:2000qz}, we assume that the Yukawa couplings are the only
source of flavor violation and (even more) that the supersymmetry breaking
parameters are universal at some fundamental scale.  Within the minimal
supergravity (mSUGRA) scenario \cite{msugra}, this scale is usually
taken to be $M_\text{GUT}$.  An alternative and arguably more natural
choice, however, would be the Planck scale, $M_\text{Pl} = G_N^{-1/2} =
1\cdot 10^{19}$~GeV.\footnote{Alternatively, one might choose the
  \emph{reduced} Planck scale, $M_\text{Pl} = \left(8\pi
    G_N\right)^{-1/2} = 2\cdot 10^{18}$~GeV, because it compensates for
  the factor $8\pi$ in the Einstein field equations.}  The reason to
take the MSSM unification scale instead is simply that while the use of
the renormalization group equations of the MSSM below $M_\text{GUT}$ is
undisputed, the analysis of the region between $M_\text{GUT}$ and
$M_\text{Pl}$ requires knowledge about the grand-unified model.  
However, the universality of the SUSY-breaking parameters is broken
by their evolution down to lower energies. Thus the choice of $M_\text{GUT}$
eliminates potentially important flavor effects.
In our analysis, we will adopt $M_\text{Pl}$ as universality scale,
and study consequences of this choice in detail.

In the standard model, fermion mixing is only measurable among the
left-handed states and described by the quark and lepton mixing
matrices, $V_\text{CKM}$ and $V_\text{PMNS}$.  Both small and large
mixing angles are realized: while those
in the quark sector are small,
two angles in $V_\text{PMNS}$ turn out to be large.  These are the neutrino solar and
atmospheric mixing angles, where the latter is close to maximal.  The
effects of $V_\text{CKM}$ and $V_\text{PMNS}$ are confined to the quark
and to the lepton sectors, respectively.  In GUTs, however, this
separation of quark and lepton sector is abrogated as quarks and leptons
are unified.  Thus their masses and mixings are related to each other.
While different patterns are possible, it is natural to expect imprints
of $V_\text{PMNS}$ on the quark sector as well.  In particular, it might
be possible to trade off small rotations of left-handed down quarks
and right-handed leptons against large mixings among right-handed down
quarks and left-handed leptons, as we will discuss below.  The mixing of
the right-handed fermions is unobservable due to the absence of right-handed
flavor-changing currents at the weak scale.  With
weak-scale supersymmetry, the mixing of the corresponding scalar
partners of quarks and leptons becomes physical.

The impact of the large atmospheric mixing angle on $B_s$ physics has
already been investigated in detail
\cite{Chang:2002mq,gut-atm-b,Jager:2003xv,cmm-bottom}.  Due to the good agreement
of the bottom-quark and tau-lepton masses at $M_\text{GUT}$, one can
adopt the predicted Yukawa unification of down quarks and charged
leptons.  In order to study $K$ and $B_d$ physics, however, one needs to go
beyond minimal models and modify the relations among the Yukawa
couplings.\footnote{These modifications were neglected in
  Ref.~\cite{Akama:2001em}, whose authors consider minimal SU(5).
  Similarly, Ref.~\cite{Barbieri:1995rs} assumes a minimal SO(10)
  model where $V_\text{CKM}$ describes all SM flavor mixing (the study
  is from 1995, i.e.\ before the large mixing angles in the lepton
  sector were established).}  Here, we can pursue two avenues: we can
either introduce additional Higgs fields in larger representations,
such as a $\text{45}_H$ in SU(5), or parameterize the modifications
via higher-dimensional operators, suppressed by powers of a more
fundamental scale \cite{Ellis:1979fg,NRO}.  We opt for the latter route
for three reasons.  One, large Higgs representations introduce a large
number of additional fields, which both yields large threshold
corrections at $M_\text{GUT}$ and makes the gauge coupling blow up
shortly above the GUT scale.  Two, the use of higher-dimensional
operators reflects the successful Yukawa unification of the third
generation; the corrections are suppressed and therefore apply mostly
to the lighter generations.  Finally, we are able to perform a more
general study as we do not rely on specific Higgs fields.

In this paper, we will study the impact of the higher-dimensional Yukawa
operators on $K-\ovl{K}$ and $B_d-\ovl{B_d}$ mixing.
A SUSY-SO(10) GUT with universal supersymmetry-breaking
parameters at the Planck scale will serve as our specific framework.
In particular, the precise measurement
of $\epsilon_K$ will enable us to tightly constrain the additional (s)quark mixing caused
by these operators.
The validity of our results for more general classes of grand-unified models will also be assessed.

%%%%%%%%%%%%%%%%%%%%%%%%%%%%%%%%%%%%%%%%%%%%%%%%%%%%%%%%%%%%%%%%%%%%%%%%
\section{Yukawa Unification and Dimension-five Operators}
\label{se:su5}
%%%%%%%%%%%%%%%%%%%%%%%%%%%%%%%%%%%%%%%%%%%%%%%%%%%%%%%%%%%%%%%%%%%%%%%%

Grand-unified theories using small Higgs representations to break the
electroweak symmetry generically predict the unification of down-quark
and charged-lepton masses \cite{Pati:1973uk,Georgi:1974sy}.\footnote{The unification of down-quark and charged-lepton masses is a prediction of the $\text{SU(4)}$ symmetry, which is present in the Pati-Salam model and respected in minimal SU(5).} Before
turning to SO(10), let us consider minimal SU(5) to bring out the central idea of this work.
Here the down-quark singlet, $d^c$, and lepton doublet, $L$, fill up the
$\ovl{\text{5}}$ representation, whereas the quark doublet, $Q$, as well as the
up-quark and the electron singlets, $u^c$ and $e^c$, are embedded in the
10.  As usual, these are left-chiral superfields; for instance, we have
the electron singlet $e_L^c$ instead of the right-handed electron $e_R$.
The adjoint Higgs field $\Sigma$ breaks SU(5) to the standard-model
group, which is then broken to $\text{SU(3)}_C \times
\text{U(1)}_\text{em}$ by a pair of quintets, $H + \ovl{H}$.

The corresponding Yukawa couplings read
\begin{align}
  \label{eq:minimal-yukawa}
  W_Y = \mathsf{Y}_1^{ij}\, \epsilon_{abcde}\, \text{10}^{ab}_i\,
  \text{10}^{cd}_j\, H^e + \mathsf{Y}_2^{ij}\, \text{10}^{ab}_i\,
  \ovl{\text{5}}_{ja}\, \ovl{H}_b \;,
\end{align}
where $a,\,b,\ldots$ denote SU(5) and $i,\,j$ flavor indices.  The second
coupling yields the unification of down-quark and charged-lepton 
Yukawa couplings $\mathsf{Y}_{d,e}$ (and thus of the corresponding masses). If $\mathsf{Y}_{d,e}$ are defined
such that the weak doublets are on the
left and the singlets on the right, we obtain
\begin{align}
  \label{eq:yukawa-unification}
  \mathsf{Y}_d = \mathsf{Y}_e^\top = \mathsf{Y_2} \;.
\end{align}
The mixings of the right-handed (left-handed) down quarks are thus identical (or, more precisely, conjugated) to those of the left-handed
(right-handed) charged leptons.

This relation works remarkably well for the third generation but not for
the lighter ones.  Thus we need to include corrections, which are
generically generated by higher-dimensional Yukawa operators, suppressed
by powers of the Planck scale, $M_\text{Pl}$ \cite{Ellis:1979fg}.  With
the given particle content, we have two operators of mass-dimension five
contributing to the down-quark and charged-lepton masses
\cite{Ellis:1979fg},
\begin{align}
  \label{eq:su5-dim5-ops}
  \mathsf{Y}_{\sigma 1}^{ij}\, \text{10}_i^{ab}\, \ovl{\text{5}}_{ja}\,
  \frac{\Sigma_b^c}{M_\text{Pl}}\, \ovl{H}_c + \mathsf{Y}_{\sigma
    2}^{ij}\, \text{10}_i^{ab}\, \ovl{\text{5}}_{jc}\,
  \frac{\Sigma_b^c}{M_\text{Pl}}\, \ovl{H}_a \;.
\end{align}
The vacuum expectation value (vev) of $\Sigma$ is proportional to
hypercharge, $\VEV{\Sigma}=\sigma \diag\left(2,2,2;-3,-3\right)$.
Hence, the second operator modifies the
relation~(\ref{eq:yukawa-unification}),
\begin{align}
  \label{eq:yukawa-unification-corr}
  \mathsf{Y}_d = \mathsf{Y}_e^\top + 5 \frac{\sigma}{M_\text{Pl}}\,
  \mathsf{Y}_{\sigma 2} \;.
\end{align}
Now we cannot diagonalize both Yukawa matrices simultaneously anymore.
In the basis where the charged leptons are diagonal, we obtain
\begin{align}
  \label{eq:yukawa-rot}
  L_d\, \mathsf{D}_d R_d^\dagger = \mathsf{D}_e + 5
  \frac{\sigma}{M_\text{Pl}}\, \widetilde{\mathsf{Y}}_{\sigma 2} \;;
\end{align}
$\mathsf{D}_i$ denote the diagonal Yukawa matrices, and $L_d$ and $R_d$
are unitary rotation matrices for the down-quark fields.  The good
agreement of the bottom and tau masses at the GUT scale indicates that
the rotation matrices $L_d$ and $R_d$ have a non-trivial 1-2 block
only,\footnote{Even if $\mathsf{Y}^{33}_{\sigma 2} \sim 1$,
  it is suppressed with respect to $\mathsf{Y}^{33}_2$ by
  $\sigma/M_\text{Pl}$.}
\begin{align}
  L_d , \, R_d \sim
  \begin{pmatrix}
    \ast & \ast & 0 \cr \ast & \ast & 0 \cr 0 & 0 & 1
  \end{pmatrix}
  .
\end{align}
Hence, the effect of the additional rotations may only be seen in
observables involving the first and second generations.
  
The effect of the dimension-five operators on proton decay has been
studied in great detail \cite{pdceay-dim5}.  In this paper, we point out
that the rotation matrix $R_d$ can be severely constrained
by the precise measurements in $K$ and $B_d$ physics.  This, in turn,
allows for a complementary study of these operators and thus enables us
to probe grand-unified models.

In the following, we will omit the indices of the
  higher-dimensional operators.  For instance, we will denote the
operators in Eq.~(\ref{eq:su5-dim5-ops})
\begin{align}
  \mathsf{Y}_{\sigma}^{ij}\, \text{10}_i\, \ovl{\text{5}}_j\,
  \frac{\Sigma}{M_\text{Pl}}\, \ovl{H} \equiv \mathsf{Y}_{\sigma
    1}^{ij}\, \text{10}_i^{ab}\, \ovl{\text{5}}_{ja}\,
  \frac{\Sigma_b^c}{M_\text{Pl}}\, \ovl{H}_c + \mathsf{Y}_{\sigma
    2}^{ij}\, \text{10}_i^{ab}\, \ovl{\text{5}}_{jc}\,
  \frac{\Sigma_b^c}{M_\text{Pl}}\, \ovl{H}_a \;.
\end{align} 
Note that these index-less operators represent all possible combinations
for the fields to form a singlet, and so $\mathsf{Y}_\sigma$ is an
\emph{effective} coupling matrix.

%%%%%%%%%%%%%%%%%%%%%%%%%%%%%%%%%%%%%%%%%%%%%%%%%%%%%%%%%%%%%%%%%%%%%%%%
\section{Framework}\label{sec:fram}
%%%%%%%%%%%%%%%%%%%%%%%%%%%%%%%%%%%%%%%%%%%%%%%%%%%%%%%%%%%%%%%%%%%%%%%%

Let us now turn to SO(10) and consider a model proposed by Chang,
Masiero, and Murayama (CMM) \cite{Chang:2002mq}.
Here the matter fields are unified in the spinor representations,
$\text{16}_i$, together with the right-handed neutrinos.  SO(10) is
broken to SU(5) by a pair of Higgs spinors, $\text{16}_H + \ovl{\text{16}}_H$.
Next, an adjoint field, $\text{45}_H$, breaks SU(5) and the electroweak
symmetry is eventually broken by a pair of fundamental Higgs fields,
$\text{10}_H$ and $\text{10}_H^\prime$.
In fact, both the SU(5) adjoint and the SU(5) singlet of
$\text{45}_H$ acquire vevs, the latter (denoted by $v_0\sim10^{17}\gev$) being an order of magnitude
larger than the former ($\sigma\sim10^{16}\gev$).

The Yukawa couplings in the CMM superpotential read
\begin{align}
  \label{eq:cmm-wy}
  W_Y & = \text{16}_i\, \mathsf{Y}_1^{ij}\, \text{16}_j\, \text{10}_H
  + \text{16}_i\, \mathsf{Y}_2^{ij}\, \text{16}_j\,
  \frac{\text{45}_H\, \text{10}_H^\prime}{M_\text{Pl}} + \text{16}_i\,
  \mathsf{Y}_N^{ij}\, \text{16}_j \frac{\ovl{\text{16}}_H
    \ovl{\text{16}}_H}{M_\text{Pl}} \ .
\end{align} 
Let us discuss the individual terms in detail.  
In the fundamental Higgs field $\text{10}_H$, only the up-type Higgs
doublet $H_u$ acquires a weak-scale vev such that the first term gives masses to the up quarks
and neutrinos only.  The masses for the down quarks and charged
leptons are then generated through the vev of the down-type Higgs
doublet $H_d$ in the second fundamental Higgs field $\text{10}_H^\prime$.
(A second Higgs field with Yukawa couplings to the SM fermions is generally needed in order to have a non-trivial flavor structure.)
They are obtained from
the second term in Eq.~(\ref{eq:cmm-wy}) which is of mass-dimension
five, in contrast to minimal SU(5).  As indicated above, this operator
actually stands for various, inequivalent effective operators with
both the SU(5)-singlet and the SU(5)-adjoint vevs of the adjoint Higgs
field $45_H$ such that the coupling matrix $\mathsf{Y}_2$ can only be
understood symbolically.  The magnitude of this second mass term is determined by the vev of the
SU(5)-singlet component, $v_0$, which contributes equally to down-quark and
charged-lepton masses.  
The strong hierarchy between the $t$ and $b,\tau$ masses then follows from the $v_0/M_\text{Pl}$ suppression factor.
The smaller SU(5)-breaking vev ($\sigma$), which is proportional
to hypercharge as in SU(5), will be important for the modification
of the light generation Yukawa couplings. 
The second term in Eq.~(\ref{eq:cmm-wy}) can be constructed in various ways, for
example by integrating out SO(10) fields at the Planck scale.  The
corresponding couplings can be symmetric or antisymmetric
\cite{so10-dim5}, resulting in an asymmetric effective coupling matrix $\mathsf{Y}_2$, as opposed to the symmetric matrices
$\mathsf{Y}_1$ and $\mathsf{Y}_N$.  Finally, the third term in
Eq.~(\ref{eq:cmm-wy}), again a higher-dimensional operator, generates
Majorana masses for the right-handed neutrinos.

We can always choose a basis where one of the Yukawa matrices in Eq.~(\ref{eq:cmm-wy}) is
diagonal.  In particular, the basis where $\mathsf{Y}_1^{ij}$ is diagonal will be referred to as the up-basis.
In the CMM model, however, one assumes that
$\mathsf{Y}_1^{ij}$ and $\mathsf{Y}_N^{ij}$ are simultaneously
diagonalizable.
This assumption is motivated by the observed values for the fermion
masses and mixings and might be a result of family symmetries.  First,
we note that the up quarks have a stronger hierarchy than the down
quarks, charged leptons, and neutrinos.  Consequently, the eigenvalues
of $\mathsf{Y}_N$ must almost have a double hierarchy compared to
$\mathsf{Y}_1$.  Then, given the Yukawa couplings in an arbitrary
basis, we expect smaller off-diagonal entries in the rotation matrices
of $\mathsf{Y}_1$ and $\mathsf{Y}_N$ than in $\mathsf{Y}_2$ because
hierarchical masses generically correspond to small mixing.
Moreover, the light neutrino mass matrix implies that, barring cancellations,
the rotations needed to diagonalize $\mathsf{Y}_1$ should be smaller than those in $V_\text{CKM}$
\cite{rot-seesaw}.  Thus, even if $\mathsf{Y}_N$ is not exactly diagonal in the up-basis,
the off-diagonal entries in its rotation matrix will be much smaller
than the entries in $V_\text{CKM}$ so that they cannot spoil the large
mixings among $d_R$ quarks generated by $V_\text{PMNS}$.

Now, with $\mathsf{Y}_1$ and $\mathsf{Y}_N$ being simultaneously
diagonal, the flavor structure is (apart from su\-per\-sym\-me\-try-breaking
terms, which we will discuss below) fully contained in the remaining
coupling, $\mathsf{Y}_2$.
Let us assume for the moment that the
relation~(\ref{eq:yukawa-unification}) is valid.  Then we can rewrite
the superpotential in the SU(5) basis as
\begin{align}
  W_Y & = \text{16}_i\, \mathsf{D}_1^{ij}\, \text{16}_j\, \text{10}_H +
  \text{16}_i \left( V_q^\ast \mathsf{D}_2 V_\ell \right)^{ij}\,
  \text{16}_j\, \frac{\text{45}_H\, \text{10}_H^\prime}{M_\text{Pl}} +
  \text{16}_i\, \mathsf{D}_N^{ij}\, \text{16}_j \frac{\ovl{\text{16}}_H
    \ovl{\text{16}}_H}{M_\text{Pl}} \ ,
\end{align}
where the second coupling is to be understood as
$\left(Q,e^c\right)^\top V_q^\ast \mathsf{D}_2 V_\ell \left(d^c,L\right)
\text{45}_H \text{10}_H^\prime / M_\text{Pl}$
(cf.~Sec.~\ref{se:su5}).  Then $V_q$ and $V_\ell$ coincide with the
quark and lepton mixing matrices, $V_\text{CKM}$ and $V_\text{PMNS}$,
up to phases.\footnote{In the up-basis, $V_\text{CKM}$ is conventionally defined as the
  matrix that rotates the left-handed down-quark mass eigenstates into
  the weak eigenbasis, while the
    inverse of $V_\text{PMNS}$ rotates the corresponding charged
    leptons.  
  The transposition between $R_d$ and $V_\ell$ in Eq.~(\ref{eq:rd}) is
  due to relation~(\ref{eq:yukawa-unification}).}
Note that the mass matrices of both down quarks and charged leptons have
a lopsided structure.

As discussed in the previous section, the
relation~(\ref{eq:yukawa-unification}) needs to be modified.  Using the
SU(5)-breaking vev of $\text{45}_H$, $\sigma$, we obtain
\begin{align}
  \label{eq:ysigma}
  \mathsf{Y}_d = \mathsf{Y}_e^\top + 5 \frac{\sigma}{v_0}\,
  \mathsf{Y}_\sigma \;,
\end{align}
in accordance with SU(5) discussed above.  Again, this notation is
symbolic, as $\mathsf{Y}_\sigma$ stems from several distinct operators.
Without these corrections, the large atmospheric mixing angle could directly be
translated to maximal mixing between the right-handed down squarks
$\tilde{b}_R$ and $\tilde{s}_R$.  Now the CKM matrix diagonalizes
$\mathsf{Y}_d\, \mathsf{Y}_d^\dagger$ whereas the PMNS matrix
diagonalizes $\mathsf{Y}_e\, \mathsf{Y}_e^\dagger$, such that we cannot
give a general relation between the contributions of the correction
operators and additional rotations.
Let us therefore make the ansatz
\begin{align}
  \label{eq:rd}
  R_d = \left(U\, V_\ell\right)^\top ,
\end{align}
i.e., the rotation of the down-quark singlet fields differs from that of
the lepton doublets by a unitary matrix $U$.  Clearly, in absence of the
correction operators, $U=\mathbbm{1}$.  As said before, the goal of this paper is to
study how much the rotations parameterized by $R_d$ differ from those of
the charged leptons, i.e.~whether a sizeable admixture of $\tilde{d}_R$
in $\tilde{s}_R^\prime$ is allowed.

As discussed above, the good bottom-tau unification implies that the
(33)-entry of $U$ should be close to one, up to a phase, and the
remaining entries of the third row and column should be small.  Thus
we parameterize $U$ as
\begin{align}
  \label{eq:m}
  U & =
  \begin{pmatrix}
    U_{11} & U_{12} & 0 \cr U_{21} & U_{22} & 0 \cr 0 & 0 & e^{i\phi_4}
  \end{pmatrix}
  =
  \begin{pmatrix}
    \cos\theta\, e^{i\phi_1} & -\sin\theta\,
    e^{i\left(\phi_1-\phi_2+\phi_3\right)} & 0 \cr \sin\theta\,
    e^{i\phi_2} & \cos\theta\, e^{i\phi_3} & 0 \cr 0 & 0 & e^{i\phi_4}
  \end{pmatrix},
\end{align}
with $\theta\in[0,\pi/2]$. For concreteness, let us assume the tribimaximal form for the leptonic
mixing matrix, corresponding to the mixing angles $\theta_{12} =
\arcsin\left(1/\sqrt{3}\right) \simeq 35^\circ$, $\theta_{13} =
0^\circ$, and $\theta_{23} = 45^\circ$.
In the up-basis, we can have $V_q$ in its standard parametrization and
thereby absorb five of the six phases.  Then we can indeed identify $V_q
= V_\text{CKM}$.  We cannot do so for $V_\ell$ since we would only move
the phases from the down-quark Yukawa matrix to the down-squark soft-breaking masses.  We therefore
choose to have $V_\ell$ with six phases; to see them explicitly, let us
write down the mixing matrix for $\theta_{13} \not= 0$,
\begin{align}
  \label{eq:mns-phases}
  V_\ell & =
  \begin{pmatrix}
 \sqrt{\frac{2}{3}} c_{13}\, e^{i\alpha_1} & \frac{1}{\sqrt{3}}
    c_{13}\, e^{i\alpha_2} & s_{13}\, e^{i\left(\delta+\alpha_3\right)}
    \\[3pt]
    e^{i\alpha_4} \left( -\frac{1}{\sqrt{6}} - \frac{1}{\sqrt{3}}
      s_{13}\, e^{-i\delta} \right) &
    e^{i\left(-\alpha_1+\alpha_2+\alpha_4\right)} \left(
      \frac{1}{\sqrt{3}} - \frac{1}{\sqrt{6}} s_{13}\, e^{-i\delta}
    \right) & \frac{1}{\sqrt{2}} c_{13}\,
    e^{i\left(-\alpha_1+\alpha_3+\alpha_4\right)}
    \\[3pt]
    e^{i\alpha_5} \left( \frac{1}{\sqrt{6}} - \frac{1}{\sqrt{3}}
      s_{13}\, e^{-i\delta} \right) &
    e^{i\left(-\alpha_1+\alpha_2+\alpha_5\right)} \left(
      -\frac{1}{\sqrt{3}} - \frac{1}{\sqrt{6}} s_{13}\, e^{-i\delta}
    \right) & \frac{1}{\sqrt{2}} c_{13}\,
    e^{i\left(-\alpha_1+\alpha_3+\alpha_5\right)}
  \end{pmatrix}
  ,
\end{align}
where $c_{13}=\cos\theta_{13}$ and $s_{13}=\sin\theta_{13}$.  In this
parametrization, we can easily identify the standard phase, $\delta$,
and then the standard form for $V_\text{PMNS}$ is given by
$V_\text{PMNS} = P_L V_\ell P_R$, where
\begin{align}
 P_L & = \diag\left( e^{-i\alpha_1},\, e^{-i\alpha_4},\,
    e^{-i\alpha_5}\right) , & P_R & = \diag\left(1,\,
    e^{i\left(\alpha_1-\alpha_2\right)},\,
    e^{i\left(\alpha_1-\alpha_3\right)} \right) .
\end{align}
If $\theta_{13}$ is indeed zero, the phase $\delta$ drops out of the
matrix (\ref{eq:mns-phases}).  This situation is familiar from the
standard model: CP violation requires $\theta_{13} \not= 0$.
Altogether, for $\theta_{13}=0$, the mixing matrix for the
right-handed down quarks in Eq.~(\ref{eq:rd}) reads
\begin{align}
  \label{eq:Rd}
\hspace*{-0.6cm}  R_d & = \frac{1}{\sqrt{6}}
  \begin{pmatrix}
 		  2 U_{11}\, e^{i\alpha_1} - U_{12}\, e^{i\alpha_4} 
 		& 2 U_{21}\,e^{i\alpha_1} - U_{22}\, e^{i\alpha_4} 
    & e^{i\left(\phi_4+\alpha_5\right)}
    \\[2pt]
    \sqrt{2}\, e^{i\alpha_2} \left( U_{11} + U_{12}\,
      e^{i\left(\alpha_4-\alpha_1\right)} \right)
    & \sqrt{2}\,e^{i\alpha_2} \left( U_{21} + U_{22}\,
      e^{i\left(\alpha_4-\alpha_1\right)} \right) 
    & -\sqrt{2}\,e^{i\left(\phi_4-\alpha_1+\alpha_2+\alpha_5\right)}
    \\[2pt]
    \sqrt{3} U_{12}\, e^{i\left(-\alpha_1+\alpha_3+\alpha_4\right)} 
    & \sqrt{3} U_{22}\, e^{i\left(-\alpha_1+\alpha_3+\alpha_4\right)}
    & \sqrt{3}\, e^{i\left(\phi_4-\alpha_1+\alpha_3+\alpha_5\right)}    
  \end{pmatrix}
\end{align}
with $U_{ij}$ as given in Eq.~(\ref{eq:m}).

Due to the absence of right-handed multiplets in the standard model,
mixing among the right-handed down quarks is unobservable.  With
supersymmetry, however, the mixing of the corresponding squarks
potentially leads to enhanced amplitudes for flavor-changing processes.
As mentioned in the Introduction, we will assume universal soft-breaking terms at the Planck
scale.
This universality, however, is no longer present at the electroweak
scale. For the scalar masses, this is due to the large Yukawa coupling
of the third generation in the renormalization group evolution (RGE), such that
\begin{align}
  \label{eq:md}
  \mathsf{M}^2_{\tilde{d}} \left(M_Z\right) & = \diag \left(
    m^2_{\tilde{d}},\ m^2_{\tilde{d}},\
    m^2_{\tilde{d}}\left(1-\Delta_{\tilde{d}}\right) \right) 
\end{align}
in the case of the $\tilde{d}_R$ soft-breaking terms.
The fast RGE between $M_\text{Pl}$ and $v_0$ allows for rather
large values of $\Delta_{\tilde{d}}$ \cite{Chang:2002mq,Jager:2003xv}. 
Now choosing the super-CKM basis where the down quarks are mass
eigenstates, this matrix is no longer diagonal; in particular, all elements of the 2-3
block are of comparable size:
\begin{align}
\begin{split}
\label{eq:mdsCKM}
\widetilde{\mathsf{M}}^2_{\tilde{d}} &= R_d^\dagger\,\mathsf{M}^2_{\tilde{d}}\, R_d
=m_{\tilde{d}}^2
\left(\begin{array}{ccc}
1-\sin^2\theta\,\Delta_{\tilde{d}}/2 &
\sin(2\theta)\,e^{-i\phi_K}\,\Delta_{\tilde{d}}/4 &
\sin\theta\,e^{-i\phi_{B_d}}\,\Delta_{\tilde{d}}/2 \\
\sin(2\theta)\,e^{i\phi_K}\,\Delta_{\tilde{d}}/4 &
1-\cos^2\theta\,\Delta_{\tilde{d}}/2 &
-\cos\theta\,e^{-i\phi_{B_s}}\,\Delta_{\tilde{d}}/2 \\
\sin\theta\,e^{i\phi_{B_d}}\,\Delta_{\tilde{d}}/2 &
-\cos\theta\,e^{i\phi_{B_s}}\,\Delta_{\tilde{d}}/2 &
1-\Delta_{\tilde{d}}/2
\end{array}\right),
\\
\phi_K & =\phi_1-\phi_2 \ , \qquad
  \phi_{B_s}=\phi_3-\phi_4+\alpha_4-\alpha_5 \ , \qquad
  \phi_{B_d}=\phi_1-\phi_2+\phi_3-\phi_4+\alpha_4-\alpha_5 \ .
\end{split}
\end{align}
This observation motivated detailed
studies of $b \to s$ transitions in supersymmetric GUT models, in
particular the decay $b \to s \gamma$ and $B_s-\ovl{B_s}$ mixing
\cite{Chang:2002mq,gut-atm-b,Jager:2003xv}.
In the following, we will study the impact of the 1-2 and 1-3 blocks,
generated by the angle $\theta$ in Eq.~(\ref{eq:m}),
on the analogous $s \to d$ and $b \to d$ transitions, focussing on $K-\ovl{K}$ and $B_d-\ovl{B_d}$ mixing.

%%%%%%%%%%%%%%%%%%%%%%%%%%%%%%%%%%%%%%%%%%%%%%%%%%%%%%%%%%%%%%%%%%%%%%%%
\section{Meson-Antimeson Mixing}\label{sec:mix}
%%%%%%%%%%%%%%%%%%%%%%%%%%%%%%%%%%%%%%%%%%%%%%%%%%%%%%%%%%%%%%%%%%%%%%%%

The oscillations of a $P^0-\ovl{P^0}$ meson system can be 
described by a Schr\"odinger-type equation,
\begin{align}
  i\frac{d}{dt}
  \begin{pmatrix}
    |P^0(t)\rangle \cr |\ovl{P^0}(t)\rangle
  \end{pmatrix}
  = \left[M^P -\frac{i}{2}\Gamma^P \right]
  \begin{pmatrix}
    |P^0(t)\rangle \cr |\ovl{P^0}(t)\rangle
  \end{pmatrix},
\end{align}
where $M^P$ and $\Gamma^P$ are two $2\times 2$ hermitian matrices
which encode the four transitions $P^0/\ovl{P^0}\to P^0/\ovl{P^0}$
via virtual and physical intermediate states, respectively.
The physical states $|P^0_1\rangle$ and $|P^0_2\rangle$
are obtained by diagonalizing $M^P -\frac{i}{2}\Gamma^P$.
The relevant quantity to study new-physics effects in $P^0-\ovl{P^0}$ mixing is the local contribution to the off-diagonal element of $M^P$:
\begin{align}
  \label{eq:m12P}
  M_{12}^P = \frac{1}{2M_P}
  \braket{P^0}{\mathcal{H}_\text{eff}^{\Delta F=2}}{\ovl{P^0}} ,
\end{align}
with $M_P$, the average meson mass $(M_{P_1}+M_{P_2})/2$.
The effective Hamiltonian $\mathcal{H}_\text{eff}^{\Delta F=2}$, which
comprises in general eight effective operators,
\begin{align}
\label{eq:Heff}
  \mathcal{H}_\text{eff}^{\Delta F=2} =
\frac{G_{F}^{2}M_{W}^{2}}{16\pi^{2}}
\sum_{i=1}^8 C^{i}_P(\mu_{P}) \, Q^{i}_P(\mu_{P}),
\end{align}
is conveniently expressed at the scale $\mu_P\sim M_P$ in the $B_d$ and $B_s$ systems, and at the scale $\mu_P\lesssim m_c$ in the kaon system.
For an extensive introduction into the formalism of $K-\ovl{K}$ and $B_{d,s}-\ovl{B}_{d,s}$ mixing, see e.g. Ref.~\cite{intro}.

One observable which is particularly well-suited to constrain the additional rotation of the $\tilde{d}_R$ and $\tilde{s}_R$ squarks in Eq.~(\ref{eq:m})~is 
\begin{align}
  \label{eq:epsilonK}
  \abs{\epsilon_K}  = \kappa_\epsilon\,\frac{\im \left(M_{12}^K\right)}{\sqrt{2}\Delta M_K} ,
\end{align}
which measures the amount of CP-violation in $K-\ovl{K}$ mixing amplitudes.
Indeed, $|\epsilon_K|$ is very small in the standard model and its experimental value, measured with high precision,
leaves only little room for new physics.
The correction factor $\kappa_\epsilon$ above parameterizes both
the small deviation of $\sin\phi_{\epsilon_K}=\Delta M_K/(\Delta M_K^2+\Delta \Gamma_K^2/4)^{1/2}$ from $1/\sqrt{2}$ and
the small contribution from the phase of the isospin-zero $K\to\pi\pi$ decay amplitude.
This factor was estimated to $\kappa_{\epsilon}=0.92\pm 0.02$ \cite{Buras:2008nn} assuming the standard model.
Its modification in the presence of new physics will not alter our analysis,
and we will ignore this complication.
The mass difference $\Delta M_{K}$ between the two eigenstates $K_L$ and $K_S$
receives both short-distance and long-distance contributions,
such that the constraint on possible new-physics effects in the short-distance~part,
\begin{align}
 \label{eq:DeltaMK}
  \left(\Delta M_{K}\right)^\text{SD} = 2\re\left(M_{12}^{K}\right),
\end{align}
is somewhat diluted among hadronic uncertainties.
Despite its precise experimental knowledge,
$\Delta M_{K}$ will thus play a minor role in our study. 

On the contrary, when new sources of CP-violation in the kaon system
  are small, two observables in the $B_d$ system will prove useful to
  gain information on the mixing angle $\theta$.  These are the mass difference,
\begin{align}
\label{eq:DeltaMd}
  \Delta M_{d} = 2 \abs{M_{12}^{B_d}} ,
\end{align}
and the coefficient of the $\sin\left(\Delta M_d\,t\right)$ term in the $B_d\to J/ \psi K_S$ time-dependent CP asymmetry,
\begin{align}
    \label{eq:SJPsiKs}
    S_{J/ \psi K_S} = \sin\left(2\beta+\phi_d^\Delta\right) \simeq \im
    \left(\frac{M_{12}^{B_d}}{\left|M_{12}^{B_d}\right|}\right) ,
    \quad \beta \equiv \arg \left[- \frac{V_{td}^{*}
        V_{tb}}{V_{cd}^{*} V_{cb}}\right] , \quad \phi_d^\Delta \equiv
    \arg\frac{M_{12}^{B_d}}{M_{12}^{B_d,\,\text{SM}}} .
  \end{align}
The phase $\phi_d^\Delta$ parameterizes CP-violating effects beyond the SM in $B_d-\ovl{B_d}$ mixing.
Here and in the following, we use the standard CKM phase convention.

Finally, we will also consider the mass difference in the $B_s$ system,
\begin{align}
\label{eq:DeltaMs}
  \Delta M_{s} = 2 \abs{M_{12}^{B_s}},
\end{align}
as well as the phase measured in
the $B_s\to J/ \psi\, \phi$ time-dependent angular distribution,
\begin{align}
    \label{eq:SJPsiPhi}
    -2\beta_s^{\rm eff} = -2\beta_s+\phi_s^\Delta \simeq \arcsin
    \left( \im \frac{M_{12}^{B_s}}{\left|M_{12}^{B_s}\right|}\right),
    \quad \beta_s \equiv -\arg\left[-\frac{V_{ts}^{*}V_{tb}}{V_{cs}^{*}V_{cb}}\right]
    , \quad \phi_s^\Delta \equiv \arg\frac{M_{12}^{B_s}}{M_{12}^{B_s,\,\text{SM}}} .
  \end{align}
In the SM, $\beta_s$ is tiny: $2\beta_s\simeq 0.04$. As long as $\phi_s^\Delta$ is not too small, we thus have $-2\beta_s^{\rm eff}\simeq\phi_s^\Delta$.
On the other hand, one also has $\phi_s\equiv\arg(-M_{12}^{B_s}/\Gamma_{12}^{B_s})\simeq\phi_s^\Delta$ \cite{Lenz:2006hd}.
In the following, we will thus identify $\phi_s=-2\beta_s^{\rm eff}$.

The current experimental values of the various observables above are reported in Tab.~\ref{tab:ExpVal}.
%%%%% Table %%%%%%%%%%%%%%%%%%%%%%%%%%%%%%%%%%%%%%%%
\begin{table}
  \centering
  \begin{tabular}{|ll|ll|}
    \hline
    $\Delta M_K^{\rm exp}=(3.483\pm0.006)\cdot 10^{-12}\mev$ & \cite{PDG08}&
    $|\epsilon_K|^{\rm exp}=(2.229\pm 0.012) \cdot 10^{-3}$ & \cite{PDG08}\\
    $\Delta M_d^{\rm exp}=(3.337\pm 0.033)\cdot10^{-10}\mev$ & \cite{PDG08} &
    $S_{J/ \psi K_S}^{\rm exp}=0.671\pm0.024$ & \cite{Barberio:2008fa}\\
    $\Delta M_s^{\rm exp}=(117.0\pm0.8)\cdot 10^{-10}\mev$ & \cite{PDG08}&
    $\phi_s^{\rm exp}=(-0.77^{\,+\,0.29}_{\,-\,0.37})\cup(-2.36^{\,+\,0.37}_{\,-\,0.29})\,\rm{rad}$ &  \cite{Barberio:2008fa}\\
    \hline
  \end{tabular}
  \caption{Current experimental values of the various $\Delta F=2$ observables considered in Eqs.~(\ref{eq:epsilonK}-\ref{eq:SJPsiPhi}).}
  \label{tab:ExpVal}
\end{table}
%%%%%%%%%%%%%%%%%%%%%%%%%%%%%%%%%%%%%%%%%%%%%%%%%%%%%%%%%%%%

\subsection{Standard-Model Contributions}
\label{sec:SM}

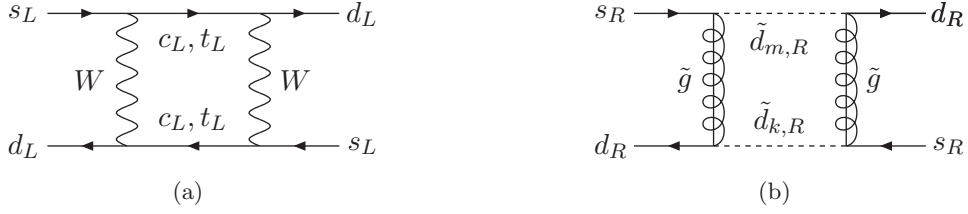
\begin{figure}
  \centering
  \subfigure[]{
    \label{fig:epsilon-k-sm}
    \centering 
    \scalebox{1}{
      \begin{picture}(150,60)(0,0)
        \ArrowLine(20,55)(50,55)    \Text(5,55)[l]{$s_L$}
        \ArrowLine(50,55)(100,55)   \Text(75,45)[]{$c_L,t_L$}
        \ArrowLine(100,55)(130,55)  \Text(145,55)[r]{$d_L$}
        \Photon(50,55)(50,5){4}{5}
        \Text(42,30)[r]{$W$}
        \Photon(100,55)(100,5){4}{5}
        \Text(108,30)[l]{$W$}
        \ArrowLine(130,5)(100,5)    \Text(145,5)[r]{$s_L$}
        \ArrowLine(100,5)(50,5)     \Text(75,15)[]{$c_L,t_L$}
        \ArrowLine(50,5)(20,5)      \Text(5,5)[l]{$d_L$}
      \end{picture}
    } }
  \hspace{50pt}
  \subfigure[]{
    \label{fig:epsilon-k-susy}
    \centering 
    \scalebox{1}{
      \begin{picture}(150,60)(0,0)
        \ArrowLine(20,55)(50,55)     \Text(5,55)[l]{$s_R$}
        \DashLine(50,55)(100,55){2}  \Text(75,45)[]{$\tilde{d}_{m,R}$} 
        \ArrowLine(100,55)(130,55)   \Text(145,55)[r]{$d_R$}
        \Gluon(50,55)(50,5){4}{5}    \Line(50,55)(50,5)
        \Text(42,30)[r]{$\tilde{g}$}
        \ArrowLine(100,55)(130,55)   \Text(145,55)[r]{$d_R$}
        \Gluon(100,55)(100,5){4}{5}  \Line(100,55)(100,5)
        \Text(108,30)[l]{$\tilde{g}$}
        \ArrowLine(130,5)(100,5)     \Text(145,5)[r]{$s_R$}
        \DashLine(100,5)(50,5){2}    \Text(75,15)[]{$\tilde{d}_{k,R}$}
        \ArrowLine(50,5)(20,5)       \Text(5,5)[l]{$d_R$}
      \end{picture}
    } }
  \caption{Dominant short-distance contributions to $M_{12}^K$ (a) in the SM; (b) in the CMM extension.}
  \label{fig:epsilon-k}
\end{figure}

In the standard model, $W$ box diagrams with virtual $t$ and/or $c$ flavors generate
the effective operators
  \begin{align}
    Q_{K}^{\rm VLL} & = \left(
      \overline{d}_{L}\gamma_{\mu}s_{L}\right) \left(
      \overline{d}_{L}\gamma^{\mu}s_{L}\right) , \qquad Q_{B_q}^{\rm
      VLL} = \left( \overline{q}_{L}\gamma_{\mu}b_{L}\right) \left(
      \overline{q}_{L}\gamma^{\mu}b_{L}\right)
  \end{align}
  for kaons (see Fig.~\ref{fig:epsilon-k-sm}) and $B_q$ ($q=s$ or
  $d$), respectively.  
The corresponding Wilson coefficients at the scale $\mu_P$ read
\begin{eqnarray}
\label{eq:WCSM}
\begin{split}
C_{K}^{\rm VLL}(\mu_K)&=4U_K(\mu_K) 
\left[
 (V_{cd}^{\ast}V_{cs})^2 \eta_1 S_0(x_c)
 +2(V_{cd}^{\ast}V_{cs})(V_{td}^{\ast}V_{ts})\eta_3 S_0(x_c,x_t)
 +(V_{td}^{\ast}V_{ts})^2\, \eta_2 S_0(x_t)
\right],\quad\\
C_{B_q}^{\rm VLL}(\mu_{B_q})&=4U_{B_q}(\mu_{B_q})(V_{tq}^{\ast}V_{tb})^2 \eta_B S_0(x_t),
\end{split}
\end{eqnarray}
where the factors
\begin{eqnarray}
\label{eq:ScaleDep}
\begin{split}
U_K(\mu)&=\left[\alpha^{(3)}_s(\mu)\right]^{-2/9}
\left[ 1+\frac{\alpha^{(3)}_s(\mu)}{4\pi}J_3\right]
\quad\text{and}\quad
U_{B_q}(\mu)&=\left[\alpha^{(5)}_s(\mu)\right]^{-6/23}
\left[ 1+\frac{\alpha^{(5)}_s(\mu)}{4\pi}J_5\right]\quad
\end{split}
\end{eqnarray}
encode the $\mu_{K},\mu_{B_q}$-dependent parts of the short-distance QCD
corrections up to next-to-leading order (NLO), while $\eta_i$ account for their $\mu_{K},\mu_{B_q}$-independent contributions \cite{Herrlich:1996vf,Buras:1990fn};
their values are given in Tab.~\ref{tab:inputs}.
The loop functions $S_0(x_q)$ and $S_0(x_c,x_t)$ are listed in the appendix.
Finally, $x_q=m_q^2/M_W^2$ and $m_q\equiv\ovl{m}_q(m_q)$ is the $\ovl{\text{MS}}$ mass.

In order to compute $M_{12}^{K,B_q}$, we still need the matrix elements of $Q_{K}^{\rm VLL}$ and $Q_{B_q}^{\rm VLL}$.
These are parameterized in terms of ``bag factors'' $B_P$, computed
at the scale $\mu=\mathcal{O}(\mu_P)$:
\begin{align}
\label{eq:BagFactors}
 \braket{P^0}{Q_{P}^{\rm VLL}(\mu)}{\ovl{P^0}}=\frac{2}{3}M^2_P F^2_P B_P(\mu),
\end{align}
where $F_P$ is the decay constant of the $P$ meson.
The scale dependences of $U_P(\mu)$ and $B_P(\mu)$ cancel each other, so that
it is convenient to define the renormalization-group-invariant parameters
$\widehat{B}_P=B_P(\mu)U_P(\mu)$.
Eqs.~(\ref{eq:WCSM}), (\ref{eq:ScaleDep}), and (\ref{eq:BagFactors}) then lead to
\begin{eqnarray}
\begin{split}
  (M_{12}^K)^\text{SM}&=\frac{G_F^2 M_W^2}{12\pi^2} M_K F_K^2 \widehat{B}_K
  \left[
  (\lambda^c_{ds})^2 \eta_1 S_0(x_c)
  +2(\lambda^c_{ds})(\lambda^t_{ds})\eta_3 S_0(x_c,x_t)
  +(\lambda^t_{ds})^2\, \eta_2 S_0(x_t)
  \right],\qquad\\
(M_{12}^{B_q})^\text{SM}&=\frac{G_{F}^{2}M_{W}^{2}}{12\pi^{2}}M_{B_q} F^2_{B_q} \widehat{B}_{B_q} (\lambda^t_{qb})^2 \eta_B S_0(x_t),
\end{split}
\end{eqnarray}
where one defines $\lambda^k_{ij}=V_{ki}^{\ast}V_{kj}$.
%%%%% Figure %%%%%%%%%%%%%%%%%%%%%%%%%%%%%%%%%%%%%%%%%%%%%
%%%%%%%%%%%%%%%%%%%%%%%%%%%%%%%%%%%%%%%%%%%%%%%%%%%%%%%%%%

\subsection{CMM Contributions}
\label{sec:CMM}
In the context of the CMM model, the dominant supersymmetric effects originate
from gluino box diagrams with virtual $\tilde{d}_R$, $\tilde{s}_R$, and $\tilde{b}_R$ flavors
due to the large mixings in Eq.~(\ref{eq:Rd}).
This gives rise to the parity-reflected operators (Fig.~\ref{fig:epsilon-k-susy})
  \begin{align}
    Q_{K}^{\rm VRR} & = \left(
      \overline{d}_{R}\gamma_{\mu}s_{R}\right) \left(
      \overline{d}_{R}\gamma^{\mu}s_{R}\right) , \qquad Q_{B_q}^{\rm
      VRR} = \left( \overline{q}_{R}\gamma_{\mu}b_{R}\right) \left(
      \overline{q}_{R}\gamma^{\mu}b_{R}\right) .
  \end{align}
The initial conditions for the Wilson coefficients at the SUSY scale
$M_S=\mathcal{O}(m_{\tilde{d}_j},m_{\tilde{g}})$ read \cite{Chang:2002mq,Jager:2003xv}
\begin{align}
\label{eq:WCCMM}
  C_{K,B_q}^{\rm CMM}(M_S) = \frac{16\pi^{2}}{G_{F}^{2}M_{W}^{2}}
  \frac{\alpha_s^2(M_S)}{2m_{\tilde{g}}^2}
  \sum_{k,m=1}^3
  (R_{d})_{mj}
  (R_{d})_{mi}^{\ast}
  (R_{d})_{kj}
  (R_{d})_{ki}^{\ast}\,
  L_0(r_m,r_k),
\end{align}
where $(i,j)=(1,2)$ in the kaon case, $(1,3)$ in the $B_d$ case, and $(2,3)$ in the $B_s$ case.
The loop function $L_0(r_m,r_k)$ is defined in the appendix,
the down-type squark mixing matrix $R_d$ was given in Eq.~(\ref{eq:Rd}),
and $r_j=m_{\tilde{d}_j}^2/m_{\tilde{g}}^2$.
Exploiting the mass degeneracy of the first two generations (see Eq.~(\ref{eq:md}))
as well as the unitarity of $R_d$, Eq.~(\ref{eq:WCCMM}) simplifies to
\begin{align}
\label{eq:WCCMM2}
  C_{K,B_q}^{\rm CMM}(M_S) = \frac{16\pi^{2}}{G_{F}^{2}M_{W}^{2}}&
  \frac{\alpha_s^2(M_S)}{2m_{\tilde{g}}^2}
  \left[ (R_{d})_{3j}(R_{d})_{3i}^{\ast} \right] ^2
  \left\lbrace L_0(r_1,r_1)-2L_0(r_1,r_3)+L_0(r_3,r_3) \right\rbrace ,\\
r_1 &= m_{\tilde{d}}^2/m_{\tilde{g}}^2, \qquad r_3 = m_{\tilde{d}}^2\left(1-\Delta_{\tilde{d}}\right)/m_{\tilde{g}}^2.
\end{align}
The RGE of the above Wilson coefficients from the scale $M_S$ down to
the scale $\mu_{K,B_q}$ is performed in two steps:
first, the leading-order matching coefficients in Eq.~(\ref{eq:WCCMM2})
are evolved down to $\mu_t=\mathcal{O}(m_t)$ by means of the leading-order RGE factor
$\eta_6=[\alpha_s^{(6)}(M_S)/\alpha_s^{(6)}(\mu_t)]^{2/7}$.
The remaining evolution, running over two orders of magnitude,
is achieved using NLO formulas -- essentially
the $U_K(\mu_K)$, $\eta_2$, $U_{B_q}(\mu_{B_q})$, and $\eta_B$ factors of Sec.~\ref{sec:SM}.
The $\mathcal{O}(\alpha_s)$ QCD corrections to the SM function $S_0(x_t)$ at the scale $\mu_t$, 
which are contained in $\eta_2$ and $\eta_B$, should be removed. Denoting them by $r=0.985$ \cite{Buras:1990fn}, we get
\begin{align}
\label{eq:Evol}
  C^{\rm CMM}_K(\mu_K) &= U_K(\mu_K)\, \eta_2\, \frac{1}{r}\, \eta_6\, C^{\rm CMM}_K(M_S),
\end{align}
and similarly for $C^{\rm CMM}_{B_q}(\mu_{B_q})$.
The cancellation of the $\mu_t$-dependence between the two parts of the evolution
is of course incomplete, yet this is a numerically small effect which can be neglected.

The bag parameters of the effective operators $Q_{K}^{\rm VRR}$ and $Q_{B_q}^{\rm VRR}$ 
are identical to those of the SM operators in Eq.~(\ref{eq:BagFactors})
such that the CMM contributions to the matrix elements $M_{12}^{P}$ finally read
\begin{eqnarray}\label{eq:M12CMM}
\begin{split}
 (M_{12}^K)^\text{CMM}&=
  \frac{\alpha_s^2(M_S)}{6m_{\tilde{g}}^2} M_K F_K^2 \widehat{B}_K\,
 \frac{e^{-2i\phi_K}\sin^2(2\theta)}{16} \, \frac{\eta_2 \eta_6}{r}\, S^{(\tilde{g})}(r_1,r_3),\qquad\\
(M_{12}^{B_d})^\text{CMM}&=
  \frac{\alpha_s^2(M_S)}{6m_{\tilde{g}}^2} M_{B_d} F_{B_d}^2 \widehat{B}_{B_d}\,
 \frac{e^{-2i\phi_{B_d}}\sin^2\theta}{4} \,  \frac{\eta_B \eta_6}{r}\, S^{(\tilde{g})}(r_1,r_3),\qquad\\
(M_{12}^{B_s})^\text{CMM}&=
  \frac{\alpha_s^2(M_S)}{6m_{\tilde{g}}^2} M_{B_s} F_{B_s}^2 \widehat{B}_{B_s}\,
 \frac{e^{-2i\phi_{B_s}}\cos^2\theta}{4} \,  \frac{\eta_B \eta_6}{r}\, S^{(\tilde{g})}(r_1,r_3),
\end{split}
\end{eqnarray}
where we explicitly display the factors $\left(R_d\right)_{3i}$
  in Eq.~(\ref{eq:WCCMM2}), $S^{(\tilde{g})}(r_1,r_3)=L_0(r_1,r_1)-2L_0(r_1,r_3)+L_0(r_3,r_3)$, and the CMM phases $\phi_{K},\ \phi_{B_d}$, and $\phi_{B_s}$ have been defined in Eq.~(\ref{eq:mdsCKM}).
Note that they fulfill the relation $\phi_{B_d}=\phi_K+\phi_{B_s}$.

\subsection{Additional Supersymmetric Contributions}
\label{sec:further}
Finally, we comment on the supersymmetric contributions
which do not exhibit the large enhancement factors characteristic of the CMM model,
namely charged-Higgs($H$)-quark and chargino($\chi$)-squark box diagrams. They do not introduce new operators, and the flavor structure of the corresponding matrix elements
is the same as in the SM,
\begin{eqnarray}
\begin{split}
 (M_{12}^K)^{H+\chi}&=\frac{G_F^2 M_W^2}{12\pi^2} M_K F_K^2 \widehat{B}_K
  \left\{2(\lambda^c_{ds})(\lambda^t_{ds})\eta^H_3 S_H(c,t)
  +(\lambda^t_{ds})^2\, \eta_2 \left[S_H(t,t)+S_{\chi}(t,t)\right]\right\},\qquad\\
(M_{12}^{B_q})^{H+\chi}&=\frac{G_{F}^{2}M_{W}^{2}}{12\pi^{2}}M_{B_q} F^2_{B_q} \widehat{B}_{B_q} (\lambda^t_{qb})^2 \eta_B \left[S_H(t,t)+S_{\chi}(t,t)\right].
\end{split}
\end{eqnarray}
The loop functions $S_H(c,t)$, $S_H(t,t)$, and $S_{\chi}(t,t)$ are given explicitly in Ref.~\cite{Buras:2000qz}. The factor $\eta^H_3=0.21$ \cite{Buras:2000qz} denotes leading-order QCD corrections to the charged-Higgs box with virtual flavors $(c,t)$. Numerically, charged-Higgs and chargino contributions are small compared to CMM effects. We checked explicitly that they can be neglected in our analysis.

%%%%%%%%%%%%%%%%%%%%%%%%%%%%%%%%%%%%%%%%%%%%%%%%%%%%%%%%%%%%%%%%%%%%%%%%
\section{Numerical Analysis}\label{sec:num}
%%%%%%%%%%%%%%%%%%%%%%%%%%%%%%%%%%%%%%%%%%%%%%%%%%%%%%%%%%%%%%%%%%%%%%%%

We are now ready to investigate the constraints of $K-\ovl{K}$ and $B_d-\ovl{B_d}$
mixing on the angle $\theta$ in the down-type squark mixing matrix $R_d$.
Since we do not expect a miraculous cancellation of the phases
$\phi_1$ and $\phi_2$, we will first focus on the case where
$\sin2\phi_K\sim\mathcal{O}(1)$ (Sec. \ref{sec:ScI}) and derive constraints on
$\theta$ from $|\epsilon_K|$ alone. We will then turn to the special case $\sin2\phi_K\sim0$
(Sec. \ref{sec:ScII}) where, as we will see, interesting constraints can still be obtained from
$\Delta M_K$, $\Delta M_d$, $S_{J/\psi K_S}$, and $\Delta M_d/\Delta M_s$.
%%%%% Table %%%%%%%%%%%%%%%%%%%%%%%%%%%%%%%%%%%%%%%%
\begin{table}
  \centering
  \begin{tabular}{|ll|ll|}
    \hline
    $\kappa_{\epsilon}=0.92\pm 0.02$ & \cite{Buras:2008nn} & $|V_{us}|=0.2246\pm0.0012$ & \cite{flavianet}\\
    $F_K=\left(156.1\pm0.8\right)$ MeV & \cite{flavianet} & $|V_{cb}|=(41.6\pm0.6)\cdot10^{-3}$ & \cite{PDG08}\\
    $\widehat{B}_K=0.75\pm0.07$ & \cite{Lubicz:2008am} & $|V_{ub}|=(3.95\pm0.35)\cdot10^{-3}$ & \cite{PDG08}\\
    $F_{B_s}\widehat{B}_{B_s}^{1/2}=(270\pm 30)$ MeV & \cite{Lubicz:2008am} & $\gamma=(70.7^{\,+\,5.7}_{\,-\,7.0})^\circ$ & [see text]\\
    \cline{3-4}
    $\xi\equiv\frac{F_{B_s}\widehat{B}_{B_s}^{1/2}}{F_{B_d}\widehat{B}_{B_d}^{1/2}}=(1.21\pm 0.04)$ & \cite{Lubicz:2008am} & $\eta_1=\left(1.32\pm 0.32\right)\left[\frac{1.30\,\textrm{GeV}}{\ovl{m}_c(m_c)}\right]^{1.1}$ &
    \cite{Herrlich:1996vf,ref23}\\
    \cline{1-2}
    \rule{0pt}{12pt}%%
    $\ovl{m}_c(m_c)=(1.266\pm 0.014)$ GeV & \cite{Allison:2008xk}&$\eta_2=0.57\pm 0.01$ & \cite{Buras:1990fn,ref23}\\
    $\ovl{m}_t(m_t)=\left(162.1\pm 1.2\right)$ GeV & \cite{Group:2008vn,Chetyrkin:2000yt}&$\eta_3=0.47\pm 0.05$ & \cite{Herrlich:1996vf,ref23} \\ 
    $\alpha_s(M_Z)=0.1176\pm 0.0020$ & \cite{PDG08}&$\eta_B=0.551\pm 0.007$ & \cite{Buras:1990fn,Buchalla:1995vs}\\ 
    \hline
  \end{tabular}
  \caption{Input parameters.}
  \label{tab:inputs}
\end{table}
%%%%%%%%%%%%%%%%%%%%%%%%%%%%%%%%%%%%%%%%%%%%%%%%%%%%%%%%%%%%

The values of the various input parameters adopted in our numerical analysis are reported in Tab.~\ref{tab:inputs}.
Inputs related to CKM elements have to be protected from new-physics impact. To this end, we determine the CKM matrix from the elements $|V_{ub}|$, $|V_{cb}|$, $|V_{us}|$, and $\delta$, the CP-phase in the standard parametrization, which equals the angle $\gamma$ of the unitarity triangle to very good accuracy. The three CKM elements are extracted from tree-level decays. We use $|V_{us}|=0.2246\pm0.0012$ \cite{flavianet}, the inclusive determination $|V_{cb}|=(41.6\pm0.6)\cdot10^{-3}$ \cite{PDG08},
and the average of inclusive and exclusive determinations $|V_{ub}|=(3.95\pm0.35)\cdot10^{-3}$ \cite{PDG08}. The angle $\gamma$ is determined via $\gamma=\pi-\alpha-\beta=\pi-\alpha^{\text{eff}}-\beta^{\text{eff}}$,
with $\beta^{\text{eff}}=\beta+\phi_d^\Delta/2=(21.1\pm0.9)^\circ$ from $S_{J/\psi K_S}$ \cite{Barberio:2008fa} and
$\alpha^{\text{eff}} =\alpha-\phi_d^\Delta/2=(88.2^{\,+\,6.1}_{\,-\,4.8})^\circ$ from $B\to\pi\pi,\ \pi\rho,\ \rho\rho$ decays \cite{Charles:2004jd}.
The dependence on the new-physics phase $\phi_d^\Delta$ cancels out in the sum $\alpha^{\text{eff}}+\beta^{\text{eff}}$, such that $\gamma=(70.7^{\,+\,5.7}_{\,-\,7.0})^\circ$ is indeed free from new-physics contamination.

No assumption is made on the squark mixing parameters $\theta$, $\phi_{K}$, $\phi_{B_d}$, and $\phi_{B_s}$ prior to the analysis of the observables in Tab.~\ref{tab:ExpVal}.
The supersymmetric parameters (in particular $m_{\tilde{g}},\ r_1$, and $r_3$, or equivalently $m_{\tilde{g}},\ m_{\tilde{d}}$, and $\Delta_{\tilde{d}}$), on the other hand, are chosen such as to satisfy the constraints coming from other observables.
The identification of viable sets of SUSY parameters is the subject of the next section.

\subsection{CMM Parameter Sets}
\label{sec:CMMsets}

In the CMM model, the large number of free SUSY parameters shrinks to six input parameters at the electroweak scale (in addition to $\theta$ and the CMM phases $\phi_{K},\ \phi_{B_d}$, and $\phi_{B_s}$). These can be chosen as the gluino mass $m_{\tilde{g}}$, the first-generation $\tilde{d}_R$ and $\tilde{u}_R$ soft masses $m_{\tilde{d}}$ and $m_{\tilde{u}}$,\footnote{The specification of both $m_{\tilde{d}}$ and $m_{\tilde{u}}$ fixes the D-term scalar mass splitting \cite{cmm-bottom,Drees:1986vd}.} the ratio of the (11)-elements of the trilinear and Yukawa couplings in the super-CKM basis $a_d^1=(A_d)_{11}/(Y_d)_{11}$, the phase of the $\mu$ parameter in the Higgs potential $\arg(\mu)$, and the ratio of the two Higgs-doublet vevs $\tan\beta$. The RGE links these CMM inputs to the remaining SUSY parameters via the assumption of universal soft-breaking parameters at the Planck scale and the intermediate $\text{SO}(10)$ and $\text{SU}(5)$ GUT relations.
Note that the similar input parameters in the CMM model and in specific SUSY scenarios without grand unification
still lead to very different phenomenologies. In such well-studied scenarios as mSUGRA or the CMSSM, the SUSY-breaking parameters are universal at $M_\text{GUT}$, as mentioned in the Introduction, leaving the universal gaugino and scalar masses, $m_{1/2}$ and $m_0$, the trilinear coupling $A$, as well as the sign of $\mu$ and $\tan\beta$ as free parameters.  In contrast to GUT models, however, these scenarios do not relate quarks and leptons to each other; the MSSM fields can be rotated independently and the large lepton mixing angles do not become visible in the quark sector.
%%%%% Figure %%%%%%%%%%%%%%%%%%%%%%%%%%%%%%%%%%%%%%%%%%%%%
\begin{figure}
  \centering
  \scalebox{1}{\epsfig{file=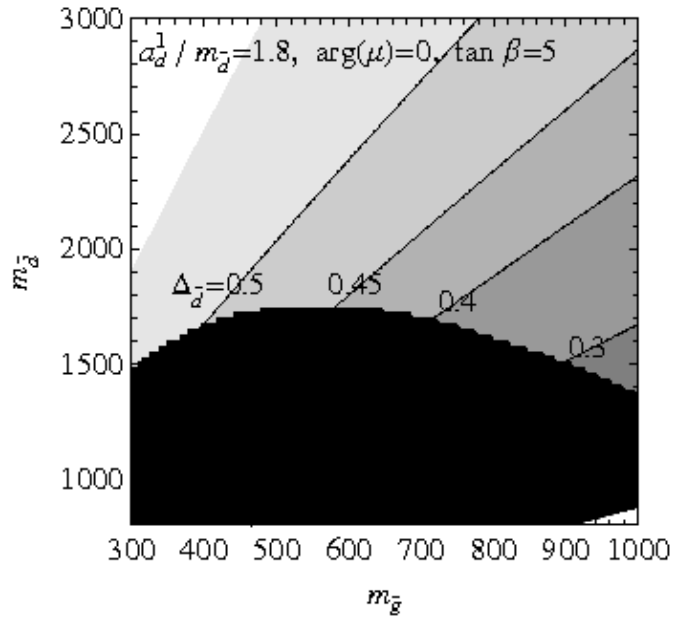}}
  \caption{Down-squark mass splitting $\Delta_{\tilde{d}}$ as a function of $m_{\tilde{g}}$ and $m_{\tilde{d}}\ [\text{GeV}]$. White: negative soft masses.
Black: excluded by lower bound on light Higgs mass.}
  \label{fig:DMsMh}
\end{figure}
%%%%%%%%%%%%%%%%%%%%%%%%%%%%%%%%%%%%%%%%%%%%%%%%%%%%%%%%%%%%

To establish benchmarks for our analysis of the down-squark mixing angle $\theta$ in $K-\ovl{K}$ and $B_d-\ovl{B}_d$ mixing, we make sure that the chosen CMM input parameters are in accord with the other observables sensitive to CMM effects, and that they respect constraints common to generic SUSY scenarios.
To this end, we make use of the \emph{Mathematica} code written by the authors of Ref.~\cite{cmm-bottom}, which implements the relations between the CMM input parameters
discussed above and the remaining SUSY parameters at the electroweak scale.
The most restrictive observable is the experimental lower bound on the mass of the lightest Higgs boson $m_h$. For small values of $\tan\beta$ it is close to the SM bound, $m_h\geq 114.4\,\text{GeV}$ \cite{Barate:2003sz}.  The main radiative corrections to the tree-level Higgs mass in the MSSM, $m_h^{\text{tree}}\leq M_Z|\cos2\beta|$, stem from (s)top loops.  For very small values of $\tan\beta\approx 3$ the large top Yukawa coupling in the RGE drives the stop mass to low values, such that the Higgs mass bound cannot be fulfilled. In our analysis we choose $\tan\beta=5$, such that the top Yukawa coupling gets smaller, but the natural hierarchy between the top and bottom Yukawa couplings, induced by $v_0/M_{\text{Pl}}$ in the CMM superpotential, is preserved.  We fix the inputs $a_d^1/m_{\tilde{d}}=1.8$ and $\arg(\mu)=0$, such that the allowed space for $m_{\tilde{g}}$ and $m_{\tilde{d}}$ around 1~TeV is large. Finally, we take $m_{\tilde{u}}=m_{\tilde{d}}$ as in Ref.~\cite{cmm-bottom}.  In Fig.~\ref{fig:DMsMh} we show the mass splitting parameter $\Delta_{\tilde{d}}$ in the $m_{\tilde{g}}-m_{\tilde{d}}$ plane for this scenario. Black regions are excluded by the Higgs mass bound.  White regions are forbidden due to negative soft mass parameters.
Additional constraints arise from processes reflecting the large atmospheric neutrino mixing angle like $\tau\to\mu\gamma$, $b\to s\gamma$, and the mass difference $\Delta M_s$; these can cut further into the low $m_{\tilde{g}}$ and $m_{\tilde{d}}$ regions.

Based on these considerations, we select three sets of CMM input parameters, given in Tab.~\ref{tab:sets}.
As said above, these parameters are defined at the electroweak scale, more precisely at $M_Z$, in Ref.~\cite{cmm-bottom}.
For consistency, we will thus set $M_S=M_Z$ (and correspondingly $\eta_6=1$, neglecting the small effect of $m_t\neq M_Z$) in our analysis of meson-antimeson mixing.
Sets 2 and 3 do satisfy the $\Delta M_{s}$ constraint for all values of $\theta$ and $\phi_{B_s}$, while Set 1 requires $|2\phi_{B_s}|$ to be between 1.2 and 2.4 radians for small $\theta$ to satisfy this constraint.
Note that especially Set 1 (with small $m_{\tilde{g}}$ and large $\Delta_{\tilde{d}}$) is chosen
such that CMM effects in $b\to s$, $b\to d$, and $s\to d$ transitions are large.

%%%%% Table %%%%%%%%%%%%%%%%%%%%%%%%%%%%%%%%%%%%%%%%%%%%%
\begin{table}
  \centering
  \begin{tabular}{|c||c|c|c||c|}
    \hline
    & $m_{\tilde{g}}$ [GeV] & $m_{\tilde{d}}$ [GeV] & 
    $\Delta_{\tilde{d}}$ & $\theta^{\text{max}}$ [$^\circ$]\\ 
    \hline
    Set 1 & 400 & 2000 & 0.52 & 0.5 \\
    Set 2 & 700 & 2000 & 0.44 & 0.9 \\
    Set 3 & 700 & 3000 & 0.51 & 0.9 \\
    \hline
  \end{tabular}
  \caption{CMM parameter sets for fixed $a_d^1/m_{\tilde{d}}=1.8,\ \arg(\mu)=0$, and $\tan\beta=5$, satisfying the constraints discussed in Sec.~\ref{sec:CMMsets}.
The last column shows the maximal mixing angle $\theta^{\text{max}}$ allowed by $|\epsilon_K|$ for $\sin2\phi_K=1$
(the symmetric solution $\theta\in[(\pi-\theta^{\rm max})/2,\pi/2]$ is excluded by $B$ physics observables, see Fig.~\ref{fig:ThetaCons2}).}
  \label{tab:sets}
\end{table}
%%%%%%%%%%%%%%%%%%%%%%%%%%%%%%%%%%%%%%%%%%%%%%%%%%%%%%%%%%%%

\subsection{Scenario I: \boldmath $\sin2\phi_K\sim\mathcal{O}(1)$ \unboldmath}
\label{sec:ScI}

\begin{figure}
  \centering
  \scalebox{0.85}{\hspace*{-0.7cm}\epsfig{file=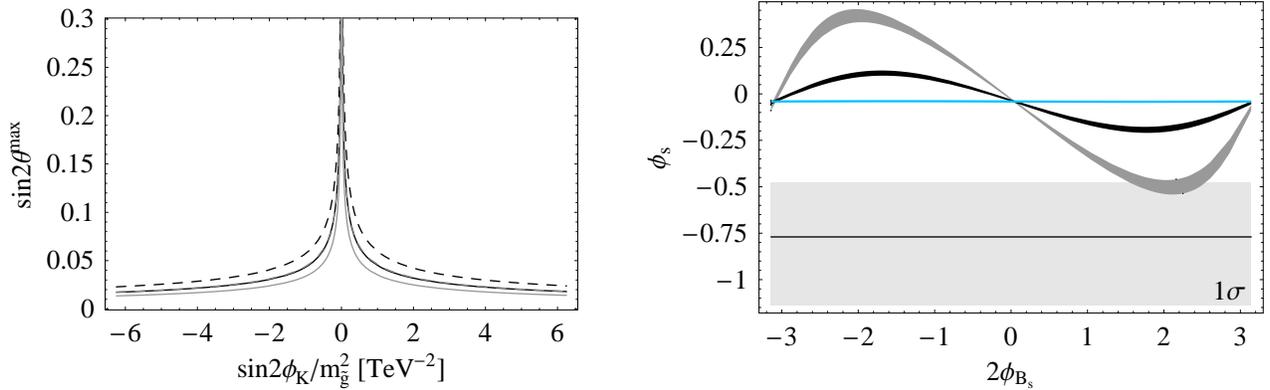}}
  \caption{Left: Constraints on $\theta$ from $|\epsilon_K|$ as a function of $\sin2\phi_K/m_{\tilde{g}}^2$ for $\Delta_{\tilde{d}}=0.44$ (black) / $0.52$ (gray) and $m_{\tilde{d}}/m_{\tilde{g}}=2.86$ (plain) / $5$ (dashed). Right: CP-violating phase $\phi_s$ as a function of the CMM phase $\phi_{B_s}$. The light gray (dark gray) curve corresponds to Set 1 (Set 2) with $\theta=0$. The SM prediction (horizontal line) is recovered for $2\phi_{B_s}=0,\pm\pi$. The broad gray band indicates the one-sigma measurement~\cite{Barberio:2008fa}.}
  \label{fig:ThetaCons1}
\end{figure}

As long as the CMM phase $\phi_K$ is not too close to zero, $|\epsilon_K|$ gives the best constraint on $\theta$.
The dependence of $\theta^{\rm max}$ on the relevant combinations of parameters, i.e., $\sin2\phi_K/m_{\tilde{g}}^2$, $m_{\tilde{d}}/m_{\tilde{g}}$, and $\Delta_{\tilde{d}}$, is summarized in Fig.~\ref{fig:ThetaCons1}-left. The plain black and dashed gray lines (which happen to be nearly superposed)
correspond to $m_{\tilde{d}}/m_{\tilde{g}}$ and $\Delta_{\tilde{d}}$ of Set 2 and Set 1, respectively, while the two other lines are obtained by interchanging $m_{\tilde{d}}/m_{\tilde{g}}$. As one can see, for $|\sin2\phi_K|/m_{\tilde{g}}^2\gtrsim1\tev^{-2}$
and typical values of the parameters $m_{\tilde{d}}/m_{\tilde{g}}$ and $\Delta_{\tilde{d}}$, $\theta^{\rm max}$ is of the order of one degree.
Fig.~\ref{fig:ThetaCons1} has been obtained treating the errors in Tab.~\ref{tab:inputs} as flat,
yet a different error treatment -- and/or inflated errors in Tab.~\ref{tab:inputs} -- would not change this picture significantly.
Fixing $\phi_K$ to $\pi/4$, the precise limits obtained for the various parameter sets defined in Sec.~\ref{sec:CMMsets} are displayed in the last column of Tab.~\ref{tab:sets}.
The small contributions in Sec.~\ref{sec:further} have no impact on these numbers.

In the $B_{d}$ and $B_{s}$ systems, the SM contributions are not as suppressed as for $\epsilon_K$.
Consequently, the smallness of $\theta^{\rm max}$ prevents any visible effect in $\Delta M_d$ and $S_{J/\psi K_S}$,
while the formulas for $\Delta M_s$ and $\phi_s$ are well approximated setting $\theta=0$.
Interestingly, sizeable CMM contributions in the $B_s$ system may be welcome to reduce the $2.2\sigma$ discrepancy between
the SM prediction for $\phi_s$ and its experimental value \cite{Barberio:2008fa}.
Within Set 1 it is possible to bring this discrepancy down to the
one-sigma level while satisfying all existing constraints, see
Fig.~\ref{fig:ThetaCons1}-right.

Finally, we briefly comment on the dependence of $\theta^{\rm max}$ on the hypothesis of tribimaximal lepton mixing.
In particular, one might expect the 23-mixing angle to be large but not $\pi/4$.  In this case,
${\rm Im}\left[ (R_{d})_{32}(R_{d})_{31}^{\ast} \right] ^2 = -\tfrac{1}{4}\sin^4\theta_{23}\sin^2\left(2\theta\right)\,
\sin(2\phi_K)$ for $\theta_{13}=0$.  Hence, for
large $\theta_{23}$, the constraints on $\theta$ do not differ much.  For a
sizeable 13-mixing angle in $V_\ell$,
$|\epsilon_K|$ gets additional contributions:
\begin{eqnarray}
\begin{split}
\Delta\left( {\rm Im}\left[ (R_{d})_{32}(R_{d})_{31}^{\ast} \right]^2 \right)=
\sin\theta_{13}\,&\sin^3\theta_{23}\,\sin\left(2\theta\right)
[    \sin(2\phi_K)\cos(2\theta)\cos(\phi_3-\phi_2+\alpha_4-\alpha_1-\delta)\\
&    \quad-\cos(2\phi_K)\sin(\phi_3-\phi_2+\alpha_4-\alpha_1-\delta)
  ] +
  \mathcal{O}\left(\sin^2\theta_{13}\right) .
\end{split}
\end{eqnarray}
No large numerical factors offset the $\sin\theta_{13}$-suppression,
so that the modified $\theta$ bounds are again as stringent as those exemplified in Fig.~\ref{fig:ThetaCons1}.

Up to now, we have taken the viewpoint of a fixed sparticle spectrum, and investigated the correlation between effects in $b\to s$ and $b,s\to d$ transitions governed by the mixing angle $\theta$. As $\theta$ turns out to be restricted to very small values, it is interesting to consider the opposite viewpoint of a fixed `natural' $\theta$ value -- say, $\sin\theta=0.5$ -- and derive the corresponding constraints on sparticle masses from $\epsilon_K$. Setting again $\phi_K$ to $\pi/4$, we find that a soft mass scale $m_{\tilde{g}}\simeq 2\,\text{TeV}$ is possible only if the ratio $m_{\tilde{d}}/m_{\tilde{g}}\simeq 1$. In such a scenario, however, the mass splitting parameter $\Delta_{\tilde{d}}$ is very small (cf. Fig.~\ref{fig:DMsMh}), such that CMM effects in other observables are negligible. For larger values of the ratio $m_{\tilde{d}}/m_{\tilde{g}}$, $\Delta_{\tilde{d}}$ increases and accordingly the constraints on $m_{\tilde{g}}$ are much more stringent (for example $m_{\tilde{g}}\gtrsim 20\,\text{TeV}$ for $m_{\tilde{d}}/m_{\tilde{g}}=2$). CMM effects in $B_d$ and $B_s$ physics are thus again killed, this time by the strong $1/m_{\tilde{g}}^2$ suppression~factor.

%%%%% Figure %%%%%%%%%%%%%%%%%%%%%%%%%%%%%%%%%%%%%%%%%%%%%
%%%%%%%%%%%%%%%%%%%%%%%%%%%%%%%%%%%%%%%%%%%%%%%%%%%%%%%%%%%%

\subsection{Scenario II: \boldmath $\sin2\phi_K\sim0$ \unboldmath}
\label{sec:ScII}

\begin{figure}
  \centering
  \scalebox{0.8}{\epsfig{file=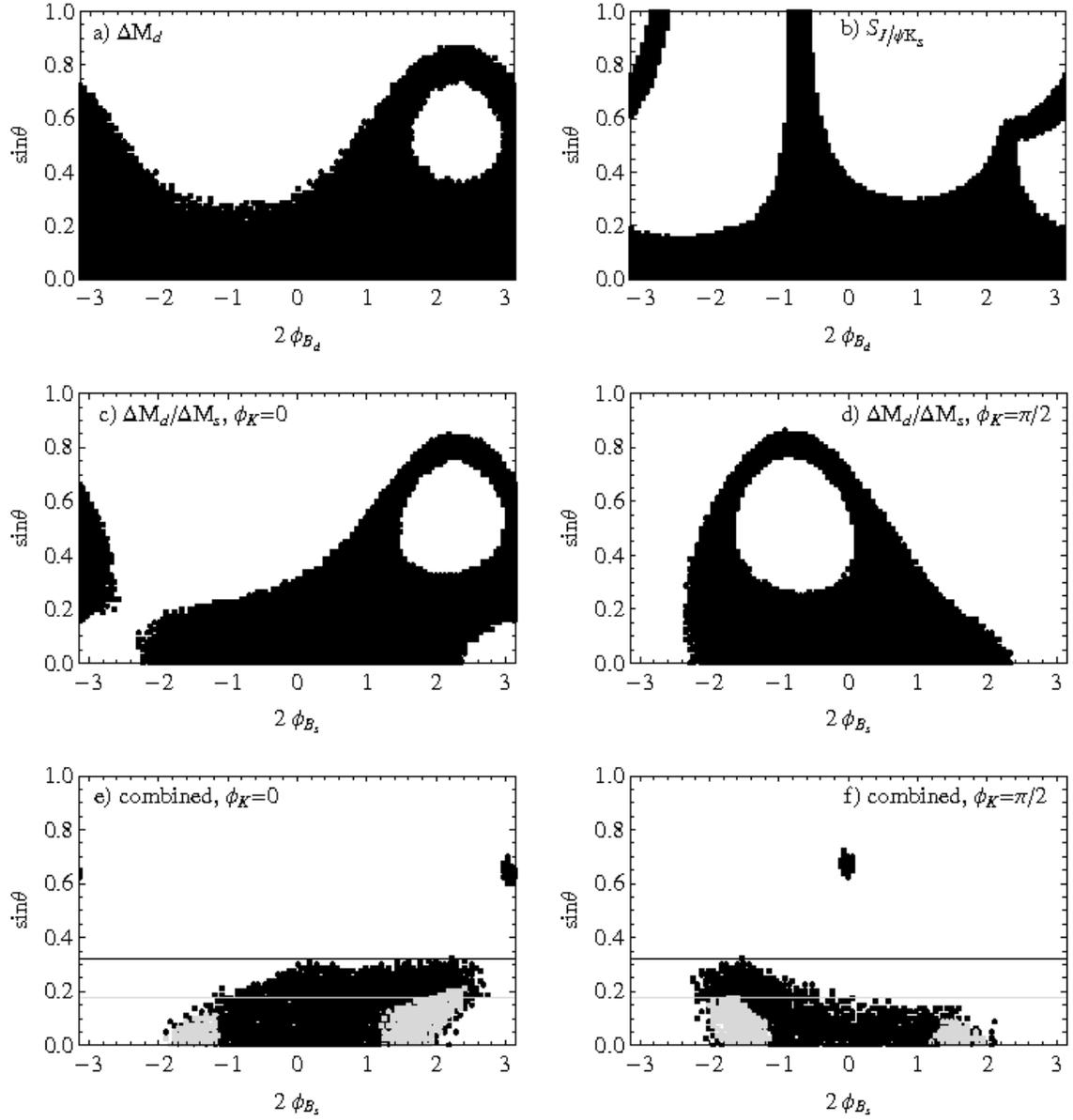}}
  \caption{Constraints on $\theta$ from $B$ physics observables. Black (gray) points indicate allowed regions in Set 2 (Set 1) parameter space.
The first four plots show individual three-sigma constraints from (a)~$\Delta M_{d}$, (b)~$S_{J/\psi K_S}$, (c)~$\Delta M_{d}/\Delta M_{s}$ setting $\phi_K=0$,
(d)~$\Delta M_{d}/\Delta M_{s}$ setting $\phi_K=\pi/2$. Plots (e) and (f) show the combined (a,b,c) and (a,b,d) constraints, respectively.
In the case of Set 1, the three-sigma constraint from $\Delta M_s$ has also been imposed, excluding points outside the $1.2\lesssim|2\phi_{B_s}|\lesssim2.4$ range (recall that Set 2 is not affected by this constraint).
Imposing further the constraint from $\phi_s$ would remove the gray points with $2\phi_{B_s}<0$
and the black points with $-1.9\lesssim2\phi_{B_s}\lesssim-1.5$ for $\sin\theta$ below $0.15$, see Fig.~\ref{fig:ThetaCons1}-right.
Finally, Set 2 (Set 1) points above the black (gray) horizontal line are excluded by~$\Delta M_K$.}
  \label{fig:ThetaCons2}
\end{figure}

If $\sin2\phi_K$ is close to zero, CMM effects cannot make their way into ${\rm Im} M^K_{12}$ anymore,
and the best constraints on $\theta$ are obtained from $\Delta M_K$ and $B$ physics observables.
As mentioned in Sec.~\ref{sec:mix}, $\Delta M_K$ is plagued by hadronic uncertainties,
so that we merely impose $|\Delta M_K^{\rm CMM}|<\Delta M_K^{\rm exp}$ to stay on the conservative side.
In this case, for $m_{\tilde{g}}\simeq700\gev$, the constraint from $\Delta M_K$ only starts to compete with that from $|\epsilon_K|$ when $|\phi_K|=\mathcal{O}(0.1^\circ)$, corresponding to $\theta^{\rm max}\simeq10^\circ-30^\circ$
(depending on the precise values of $\Delta_{\tilde{d}}$ and $m_{\tilde{d}}/m_{\tilde{g}}$).
The constraints from $\Delta M_d$, $S_{J/\psi K_S}$, and $\Delta M_d/\Delta M_s$ are in general better,
as we illustrate in Fig.~\ref{fig:ThetaCons2} for Set 1 and Set 2.
Note that the constraint from $\Delta M_d/\Delta M_s$ depends on both $\phi_{B_s}$ and $\phi_{B_d}=\phi_{K}+\phi_{B_s}$.
The plots shown in Fig.~\ref{fig:ThetaCons2} correspond to $\phi_K=0$ and $\phi_K=\pi/2$.
Other $\phi_K$ values lead to different plots, with however the same general appearance, in particular
the exclusion of small $\theta$ angles for some specific $\phi_{B_s}$ values.
For these specific values, the tight bounds on $\theta$ derived in Sec.~\ref{sec:ScI} are thus even surpassed.

As mentioned previously, $\phi_s$ can cut further into the parameter space, especially for negative $\phi_{B_s}$ values,
see Fig.~\ref{fig:ThetaCons1}-right. However, this does not change the typical value of $\theta^{\rm max}$ obtained from $B$ physics observables, which is of ten or a few tens of degrees.

%%%%% Figure %%%%%%%%%%%%%%%%%%%%%%%%%%%%%%%%%%%%%%%%%%%%%
%%%%%%%%%%%%%%%%%%%%%%%%%%%%%%%%%%%%%%%%%%%%%%%%%%%%%%%%%%%%

\subsection{Closing the Unitarity Triangle}
\label{sec:closingUT}
Recently, several studies pointed out a possible tension in the SM between the value of $\sin2\beta$ predicted from $|\epsilon_K|$ and $\Delta M_s/\Delta M_d$, and its direct measurement from $S_{J/\psi K_S}$ \cite{Lunghi:2008aa,Buras:2008nn,Buras:2009pj,Lunghi:2009sm}. In this section, we illustrate how CMM effects can remove this tension, and simultaneously account for a sizeable CP-violating phase in the $B_s$ system. %(though, as shown in Fig.~\ref{fig:ThetaCons1}-right, it is not always possible to reach the one-sigma experimental band for this phase).

Due to the particular sensitivity of $|\epsilon_K|$ to new-physics effects, either $\theta$ or $\phi_K$ must be very small. We will thus consider the two limits $\theta=0$ and $\phi_K=0$. For each case, we will compare the value of $\sin2\beta$ extracted from $S_{J/\psi K_S}$ with its determination from $|V_{us}|$, $|V_{cb}|$, $|\epsilon_K|$, and $\Delta M_{s}/\Delta M_{d}$, obtained inverting the following expressions with respect to $\sin2\beta$ and $R_t$:
\begin{eqnarray}
\begin{split}
|\epsilon_K|
&=\kappa_{\epsilon} \frac{M_K F_K^2 \widehat{B}_K}{12\sqrt{2}\Delta M_K} \times
\Biggl\lbrace \frac{G_F^2M_W^2}{\pi^2} |V_{cb}|^2|V_{us}|^2\ 
     \biggl[ |V_{cb}|^2 R_t^2 \sin(2\beta) \eta_2S_0(x_t)  \\
   & + 2R_t\sin\beta \left( \eta_3S_0(x_c,x_t)-\eta_1S_0(x_c) \right) \biggr]
- \frac{\alpha_s^2(M_S)}{8m_{\tilde{g}}^2}\sin(2\phi_K)\sin^2(2\theta)\frac{\eta_2 \eta_6}{r}\, S^{(\tilde{g})}(r_1,r_3) \Biggr\rbrace,
\end{split}
\end{eqnarray}
\begin{eqnarray}
\begin{split}
\frac{\Delta M_{s}}{\Delta M_{d}}
&= \xi^2\frac{M_{B_s}}{M_{B_d}}
\frac{\sqrt{\left(  k_1 + X \cos2\phi_{B_s}\cos^2\theta \right)^2 
          + \left( -2 k_2 R_t \sin\beta |V_{us}|^2 - X \sin2\phi_{B_s}\cos^2\theta \right)^2    }}
{\sqrt{\left( R_t^2 \cos2\beta |V_{us}|^2  + X \cos2\phi_{B_d}\sin^2\theta \right)^2
          + \left( R_t^2 \sin2\beta |V_{us}|^2  - X \sin2\phi_{B_d}\sin^2\theta \right)^2}}\, .
\end{split}
\end{eqnarray}
Here $k_1=1+|V_{us}|^2(1-2R_t\cos\beta)$, $k_2=1+|V_{us}|^2(1-R_t\cos\beta)$, 
\begin{eqnarray}
\begin{split}
X=\frac{ \pi^2 \alpha_s^2(M_S) \,\eta_6\, S^{(\tilde{g})}(r_1,r_3) }{ 2 |V_{cb}|^2 G_F^2 M_W^2 m^2_{\tilde{g}} \,r\, S_0(x_t) },
\end{split}
\end{eqnarray}
and $R_t=|V_{td}V_{tb}^{\ast}|/|V_{cd}V_{cb}^{\ast}|$ is a side of the unitarity triangle (UT).
The above expressions hold to 0.5\% accuracy.
In the SM, this leads to $\sin(2\beta^{\,\epsilon_K})=0.81^{\,+0.11}_{\,-0.09}$ with the inclusive $|V_{cb}|$ determination of Tab.~\ref{tab:inputs},
and to $\sin(2\beta^{\,\epsilon_K})=0.98^{\,+0.02}_{\,-0.11}$
if the exclusive determination from $B\to D^* \ell\nu$ decays, $|V_{cb}|^{\rm excl}=(38.8\pm1.1)\cdot10^{-3}$ \cite{Barberio:2008fa},
is used instead. Note that $|V_{cb}|^{\rm incl}$ does not lead to any significant deviation with respect to $S_{J/\psi K_S}^{\rm exp}$,
while a tension is indeed observed with the smaller value $|V_{cb}|^{\rm excl}$.
In order to illustrate how CMM effects can compensate for a low $|V_{cb}|$ input in UT analyses, we will adopt the averaged value
of Ref.~\cite{Lunghi:2009sm}, $|V_{cb}|^{LS}=(41.0\pm0.63)\cdot10^{-3}$.
In the following, we use the CMM input parameters of Set~1.
All errors are treated as gaussian.
%%%%% Figure %%%%%%%%%%%%%%%%%%%%%%%%%%%%%%%%%%%%%%%%%%%%%
\begin{figure}
  \centering
  \begin{minipage}{0.5\linewidth}
    {
      \scalebox{0.8}{\epsfig{file=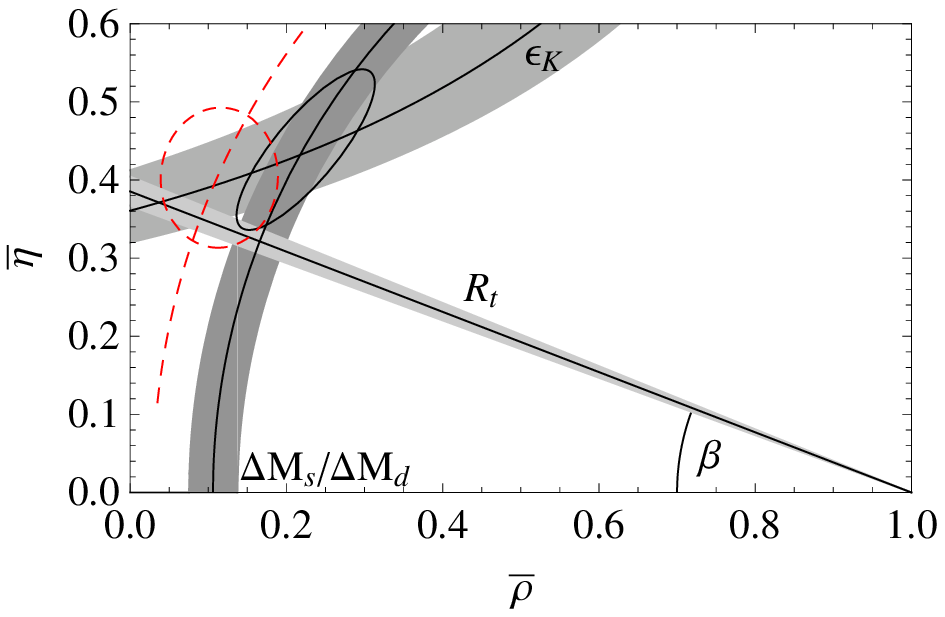}}
    }%
  \end{minipage}
  \begin{minipage}{0.45\linewidth}
    {
      \scalebox{0.8}{\epsfig{file=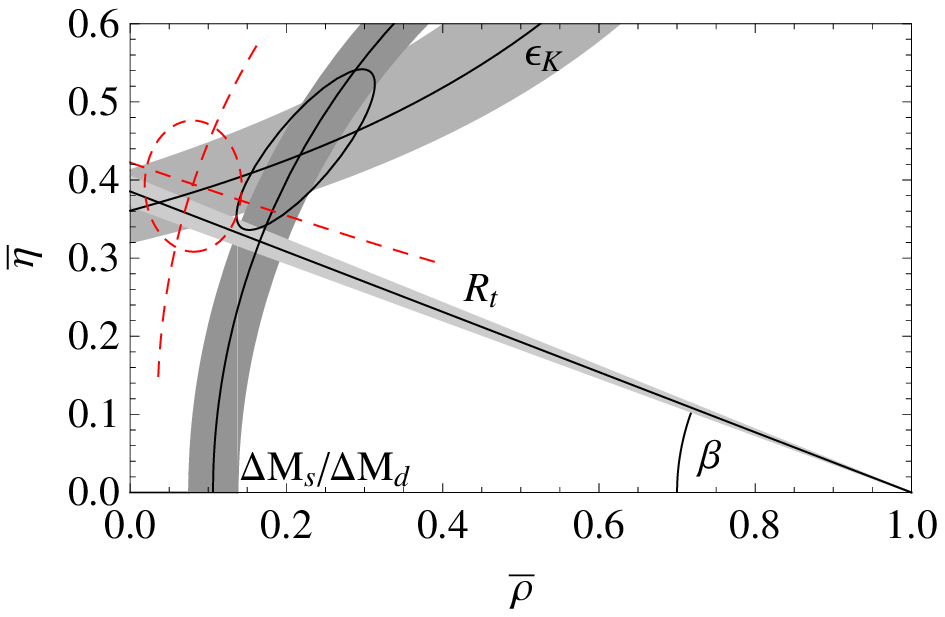}}
    }%
  \end{minipage}
  \addtolength{\abovecaptionskip}{-5pt}
  \caption{One-sigma constraints on the UT from $S_{J/\psi K_S}$ (light gray), $|\epsilon_K|$ (gray), and $\Delta M_s/\Delta M_d$ (dark gray) in the SM. The one-sigma region
determined from $|V_{us}|$, $|V_{cb}|$, $|\epsilon_K|$, and $\Delta M_{s}/\Delta M_{d}$ assuming the SM is shown in black, and its shift due to CMM effects is indicated in dashed red. Left: Scenario I, $\theta=0,\phi_{B_s}=0.7$. Right: Scenario II, $\theta=0.1,\phi_{B_s}=\phi_{B_d}=0.7$. CMM inputs: Set 1.}
  \label{fig:ut} 
\end{figure}
%%%%%%%%%%%%%%%%%%%%%%%%%%%%%%%%%%%%%%%%%%%%%%%%%%%%%%%%%%%%

\paragraph{\boldmath $\theta=0$: CMM effects in $R_t$ \unboldmath}
\ Since for $\theta=0$ there are no effects in $K$ and $B_d$ mixing, CMM contributions enter the UT only via $\Delta M_{s}$.  From Fig.~\ref{fig:ut}-left, one sees that $R_t$ has to increase in order to close the UT. This requires a CP-violating phase $2\phi_{B_s}\in [1.2,1.8]$, taking into account the three-sigma constraints on $\phi_{B_s}$ from $\Delta M_s$ and $\phi_s$. The dashed red curve shows $R_t$ for $\phi_{B_s}=0.7$, such that the UT determined from $|\epsilon_K|$ and $\Delta M_{s}/\Delta M_{d}$ agrees with the $\sin2\beta$ measurement from $S_{J/\psi K_S}$.

\paragraph{\boldmath $\phi_K=0$, $\theta=0.1$: CMM effects in $R_t$ and $\beta$ \unboldmath}
\ In this second case, CMM effects affect both $\Delta M_s/\Delta M_d$ and $S_{J/\psi K_S}$.  For a fixed angle $\theta$, the UT can be closed by adapting the CMM phase $\phi_{B_s}=\phi_{B_d}$. The resulting apex of the UT is shown by the intersection of the dashed red lines in Fig.~\ref{fig:ut}-right for $\theta=0.1$ and $\phi_{B_s}=0.7$.  For any value of $\theta$ allowed by the constraints from $B_d$ and $B_s$ observables in Sec.~\ref{sec:ScII}, one can find a phase $\phi_{B_s}$ to close the UT.

\vspace{3mm}
Deviations from these two limit cases, i.e., small but nonzero $\theta$ or $\phi_K$ values,
rapidly generate CMM effects in $|\epsilon_K|$ as well (Fig.~\ref{fig:ThetaCons1}-left).
These can lower the band from the $|\epsilon_K|$ constraint in the $(\ovl{\rho},\ovl{\eta})$ plane,
directly making up for the low $|V_{cb}|$ input value.

%%%%%%%%%%%%%%%%%%%%%%%%%%%%%%%%%%%%%%%%%%%%%%%%%%%%%%%%%%%%%%%%%%%%%%%%
\section{Conclusions}
%%%%%%%%%%%%%%%%%%%%%%%%%%%%%%%%%%%%%%%%%%%%%%%%%%%%%%%%%%%%%%%%%%%%%%%%
Grand-unified theories introduce relations among quark and lepton masses and mixings.
Motivated by the large atmospheric mixing
  angle in the neutrino sector, several studies focussed on the
  consequences of the SU(5) Yukawa relation
  $\mathsf{Y}_{d}=\mathsf{Y}_{e}^\top$ in $b \to s$ transitions.  In
  this work, we considered corrections to this relation which are
  essential to account for the observed light quark and lepton masses.
  In particular, we investigated the effects on $s \to d$ and $b \to
  d$ transitions of the additional rotation of the $d_R$ and $s_R$
  quarks.  This deviation with respect to the PMNS matrix, denoted by $U$, can be
  parameterized by an additional mixing angle $\theta$ (see
  Eqs.~(\ref{eq:rd},\ref{eq:m})).
  
In our analysis, we focussed on models with small Higgs
representations; a modified version of the CMM model served as our
specific scenario.  In this setup, the differences between the down-quark and charged-lepton masses are naturally explained by
dimension-five Yukawa operators.
The associated supplementary rotation $\theta$ was constrained
from $K-\ovl{K}$ and $B_d-\ovl{B_d}$ mixing observables.
In particular, we found that, in the absence of fortuitous cancellations
among the new phases in the matrix $U$, $|\epsilon_K|$ sets a stringent bound on $\theta$, $\theta^{\rm max}\sim\mathcal{O}(1^\circ)$.
Consequently, in the basis where the charged-lepton Yukawa couplings are diagonal, the matrix
$ \mathsf{D}_e \widetilde{\mathsf{Y}}_{\sigma} + \widetilde{\mathsf{Y}}_{\sigma}^\dagger \mathsf{D}_e
+ 5 \frac{\sigma}{v_0}\widetilde{\mathsf{Y}}_{\sigma}^\dagger \widetilde{\mathsf{Y}}_{\sigma}$
(in the notations of Eqs.~(\ref{eq:yukawa-rot},\ref{eq:ysigma}))
must be diagonal as well. Barring cancellations, this implies that
the flavor structure of the couplings which modify the Yukawa unification must be similar to that of the initial terms.
In other words, in the corrected relation $\mathsf{Y}_d = \mathsf{Y}_e^\top + 5 \frac{\sigma}{v_0}\,\mathsf{Y}_\sigma$
(Eq.~(\ref{eq:ysigma})),
the three matrices $\mathsf{Y}_\sigma,\ \mathsf{Y}_d$, and $\mathsf{Y}_e^\top$ must be essentially aligned.
Constraints from $B$ physics observables ($\Delta M_d$, $S_{J/\psi K_S}$, and $\Delta M_s/\Delta M_d$) were also analyzed,
and shown to imply the looser bound $\theta^{\rm max}\sim\mathcal{O}(10^\circ)$.

While we have worked out this analysis for a specific GUT
model, our results hold in general for models with small Higgs representations:
large effects of the neutrino mixing angles on $b_R\to s_R$ transitions
lead to large effects in $b_R\to d_R$ and $s_R\to d_R$ transitions
for natural values of the parameters, once the mass relations for the light quarks and leptons are corrected.
An efficient mechanism is naturally needed to render the mixing among right-handed $d$-quarks visible.
In the CMM model, this mechanism is provided by the fast SO(10) running of the ${\tilde d}_R$ soft
mass matrix, which generates the large universality breaking $\Delta_{\tilde d}$ at the electroweak scale.
Of course, other GUT scenarios could include additional sources of flavor and CP violation
inducing effects in $|\epsilon_K|$.
These could soften the constraints on $\theta$. Yet they would have to be
fairly fine-tuned to cancel the potentially large impact of the
  corrections from the $d_R$ rotation matrix $R_d$
  (Eq.~(\ref{eq:rd})).

Interestingly, the correction operators which are of importance
  for proton decay but contribute equally to the fermion masses ought
  to have a different flavor structure in order to be in agreement
  with the experimental limit \cite{pdceay-dim5}.  Both types of operators are
  generically present in GUTs.  Hence, our analysis is an important
  step in establishing a consistent grand-unified model.

Finally, we also considered the possible tension between the value of $\sin2\beta$ predicted from $|\epsilon_K|$ and $\Delta M_s/\Delta M_d$ in the SM and its direct measurement from $S_{J/\psi K_S}$, raised by the authors of Refs.~\cite{Buras:2008nn,Lunghi:2008aa,Buras:2009pj,Lunghi:2009sm}.
We illustrated how CMM effects can remove this tension, and simultaneously reduce the $2.2\sigma$ discrepancy observed recently in the $B_s-\ovl{B_s}$ mixing phase.

%%%%%%%%%%%%%%%%%%%%%%%%%%%%%%%%%%%%%%%%%%%%%%%%%%%%%%%%%%%%%%%%%%%%%%%%
\subsubsection*{Acknowledgements}
We are grateful to Ulrich Nierste for initiating the project and to Waldemar Martens for his help with the \emph{Mathematica} code of Ref.~\cite{cmm-bottom}. We also thank Christopher Smith and Scott Willenbrock  for useful discussions. This work is supported by the DFG grant No. NI 1105/1-1, by the DFG-SFB/TR9, by the Graduiertenkolleg ``Hochenergiephysik und Astroteilchenphysik'', and by the EU contract No. MRTN-CT-2006-035482 (FlaviaNet).

%%%%%%%%%%%%%%%%%%%%%%%%%%%%%%%%%%%%%%%%%%%%%%%%%%%%%%%%%%%%%%%%%%%%%%%%
\begin{appendix}
\section*{Appendix: Loop Functions}\label{sect:loops}
\begin{align}
  \label{eq:loop}
	S_0(x_c)&=x_c,
\\[3mm]
	S_0(x_t)&=\frac{4x_t-11x_t^2+x_t^3}{4(1-x_t)^{2}}-\frac{3x_t^3\log (x_t)}{2(1-x_t)^{3}},
\\[1mm]
	S_0(x_c,x_t)&=x_c \left[ \log\frac{x_t}{x_c}-\frac{3x_t}{4(1-x_t)}-\frac{3x_t^2\log x_t}{4(1-x_t)^{2}} \right],
\\[0.5mm]
  F(x,y) & = -\frac{1}{(x-1)(y-1)} - \frac{1}{x-y} \left[\frac{x\ln x}{(x-1)^2} - \frac{y\ln y}{(y-1)^2}\right] , 
\\
  G(x,y) & = \frac{1}{(x-1)(y-1)} + \frac{1}{x-y} \left[\frac{x^2\ln x}{(x-1)^2} - \frac{y^2\ln y}{(y-1)^2}\right] ,
\\[1mm]
  L_0(x,y) & =\frac{11}{18}G(x,y)-\frac{2}{9}F(x,y),
\\[3mm]
S^{(\tilde{g})}(x,y) &=L_0(x,x)-2L_0(x,y)+L_0(y,y).
\end{align} 
\end{appendix}

%%%%%%%%%%%%%%%%%%%%%%%%%%%%%%%%%%%%%%%%%%%%%%%%%%%%%%%%%%%%%%%%%%%%%
{\small%%
  
}%

\end{document}